\documentclass[twocolumn]{aastex63}
\pdfoutput=1 
\usepackage{amsmath,amstext,amssymb,mathtools,graphicx,color}

\usepackage{graphicx}	
\usepackage{amsmath}	
\usepackage{amssymb}	
\usepackage{xcolor}
\usepackage{lipsum}
\usepackage{hyperref}
\usepackage[flushleft]{threeparttable}
\usepackage{lineno}
\usepackage{comment}

\newcommand{\C}[1]{\textcolor{red}{#1}}
\newcommand{\G}[1]{\textcolor{orange}{#1}}

\begin{document}

\title{The 157-month {\textit Swift}-BAT All-Sky Hard X-Ray Survey}

\correspondingauthor{Amy Lien}
\email{alien@ut.edu}

\author[0000-0002-7851-9756]{Amy Y. Lien}
\affiliation{University of Tampa, Department of Physics and Astronomy, 401 W. Kennedy Blvd, Tampa, FL 33606, USA}
\affiliation{Center for Research and Exploration in Space Science and Technology (CRESST) and NASA Goddard Space Flight Center, Greenbelt, MD 20771, USA}
\affiliation{Department of Physics, University of Maryland, Baltimore County, 1000 Hilltop Circle, Baltimore, MD 21250, USA}
\author{Hans Krimm}
\affiliation{National Science Foundation 2415 Eisenhower Ave., Alexandria, VA 22314}
\author{Craig Markwardt}
\affiliation{Astrophysics Science Division, NASA Goddard Space Flight Center,Greenbelt, MD 20771, USA}
\author{Kyuseok Oh}
\affiliation{Korea Astronomy \& Space Science institute, 776, Daedeokdae-ro, Yuseong-gu, Daejeon 34055, Republic of Korea}
\author[0000-0002-8472-3649]{Lea Marcotulli}
\altaffiliation{NHFP Einstein Fellow}
\affiliation{Department of Physics \& Yale Center for Astronomy \& Astrophysics, New Haven, CT 06511, USA}
\author{Richard Mushotzky}
\affiliation{Department of Astronomy, University of Maryland, College Park, MD 20742, USA}
\affiliation{Joint Space-Science Institute, University of Maryland, College Park, MD 20742,
USA}
\author{Nicholas R. Collins}
\affiliation{Telophase Corporation at NASA’s Goddard Space Flight Center, Code 667, Greenbelt, MD 20771, USA}
\author{Scott Barthelmy}
\affiliation{Astrophysics Science Division, NASA Goddard Space Flight Center,Greenbelt, MD 20771, USA}
\author{Wayne H. Baumgartner}
\affiliation{NASA Marshall Space Flight Center, Huntsville, AL 35812, USA}
\author[0000-0003-1673-970X]{S. Bradley Cenko}
\affiliation{Astrophysics Science Division, NASA Goddard Space Flight Center,Greenbelt, MD 20771, USA}
\affiliation{Joint Space-Science Institute, University of Maryland, College Park, MD 20742, USA}
\author{Michael Koss}
\affiliation{Eureka Scientific, 2452 Delmer Street Suite 100, Oakland, CA 94602-3017, USA}
\author{Sibasish Laha}
\affiliation{Center for Space Science and Technology, University of Maryland Baltimore County, 1000 Hilltop Circle, Baltimore, MD 21250, USA.}
\affiliation{Astrophysics Science Division, NASA Goddard Space Flight Center,Greenbelt, MD 20771, USA}
\author{Takanori Sakamoto}
\affiliation{Department of Physical Sciences, Aoyama Gakuin University, 5-10-1 Fuchinobe, Chuo-ku,
Sagamihara, Kanagawa 252-5258, Japan}
\author[0000-0001-7128-0802]{David Palmer}
\affiliation{Los Alamos National Laboratory, Los Alamos, NM 87544, USA}
\affiliation{New Mexico Consortium, Los Alamos, NM, 87544, USA}
\author[0000-0002-4299-2517]{Tyler Parsotan}
\affiliation{Astrophysics Science Division, NASA Goddard Space Flight Center,Greenbelt, MD 20771, USA}


\begin{abstract}
The Burst Alert Telescope (BAT) onboard the Neil Gehrels {\it Swift} observatory has been serving as a survey instrument for the hard X-ray sky, and has detected thousands of X-ray sources (e.g., AGNs, X-ray binaries, etc). BAT monitors these X-ray sources and follows their light curves on time scales from minutes to years. In addition, BAT discovers hundreds of new X-ray sources in survey images stacked throughout the mission lifetime. We present the updated BAT survey catalog since the last published BAT 105 month survey catalog \citep{Oh18} with additional of 4.5 years of data until December 2017. Data since 2007 are reprocessed to include updated instrumental calibration.
Analysis in this study shows that additional systematic noise can be seen in the 157-month mosaic images, resulting in decreases in the expected improvement in sensitivity and the number of new detections. The BAT 157-month survey reaches a sensitivity of $8.83 \times 10^{-12} \rm \ erg \ s^{-1} \ cm^{-2}$ for 90\% of the sky and $6.44 \times 10^{-12} \rm \ erg \ s^{-1} \ cm^{-2}$ for 10\% of the sky. This catalog includes spectra, monthly and snapshot light curves in eight energy bands (14-20, 20-24, 24-35, 35-50, 50-75, 75-100, 100-150, and 150-195 keV) for 1888 sources, including 256 new detections above the detection threshold of $4.8\sigma$. The light curves, spectra, and tables that summarize the information of the detected-sources are available in the online journal and in the catalog web page \url{https://swift.gsfc.nasa.gov/results/bs157mon/}.
\end{abstract}

\section{Introduction}
\label{sect:intro}
The Burst Alert Telescope (BAT) onbaord the Neil Gehrels {\it Swift} Observatory was launched in November 20, 2004, and has been observing the hard X-ray sky ($\sim 14-195$ keV). BAT is a coded-mask instrument that is able to provide decent localization of hard X-ray sources ($\sim 3$ to $\sim 12$ arcmin for sources depends on the detection significance) while maintaining a large field of view ($\sim 1.4$ sr; half coded). BAT observes the entire sky in a mostly-uniform manner, and collects continuous survey data that are binned in $\sim 300$ s. As the mission time increases, the survey data provides a long-term monitoring of hard X-ray sources, and offers mosaic images with mission-long exposure time to detect faint sources in the hard X-ray sky.

We have published five {\it Swift}/BAT all-sky hard X-ray survey catalogs for 3-month \citep{Markwardt05}, 9-month \citep{Tueller08}, 22-month \citep{Tueller10}, 70-month \citep{Baumgartner13}, and 105-month \citep{Oh18} of data. In this 157-month survey catalog, we include another 4.5 years of data until December, 2017. Throughout the paper, we refer to the previous BAT catalog as the X-month survey catalog, for example, the 105-month survey catalog for \citet{Oh18}.

All sky X-ray surveys and the catalogs made from them have been one of the major drivers of research in X-ray astronomy from the early days \citep{Giacconi72, Piccinotti81, Voges99} providing the fundamental database from which targeted studies can be pursued, revealing new types of objects and measuring the X-ray luminosity function of many classes of sources. In this regard, the hard ($E>15$ keV) X-ray band has been one of the last frontiers of sensitive all sky surveys. The first such work \citep{Levine84} had only 44 sources. A major leap in the all-sky X-ray sensitivity was achieved by the hard X-ray survey conducted by {\it Swift}/BAT and {\it INTEGRAL}/IBIS\citep{Markwardt05, Beckmann06, Krivonos10, Tueller10, Baumgartner13, Oh18, Krivonos22_INTEGRAL}. 

The {\it INTEGRAL}/IBIS is also a coded-mask instrument and has been performing hard X-ray all-sky surveys in similar energy range ($17 - 290$ keV) to BAT since its launch in October, 2002. It has a better angular resolution than BAT ($\sim 12$ arcmin compares to the $19.5$ arcmin FWHM PSF for BAT), and has a narrower field of view \citep[$30^{\circ} \times 30^{\circ}$ partially coded FOV at zero-response and $9^{\circ} \times 9^{\circ}$ fully-coded;][]{Ubertini03, Lebrun03, Goldwurm03, Revnivtsev04}. {\it INTEGRAL}/IBIS focuses on performing sky survey near the Galactic plane and achieves longer exposure around those areas, while BAT performs a nearly uniform all-sky survey and provides longer exposures in regions away from the Galactic plane. Therefore, the {\it INTEGRAL}/IBIS and BAT catalogs offers complementary sky surveys. 

Over the past two decades, the all-sky sensitivity in the hard X-ray has improved by a factor of $\sim 30$, and the number of sources has increased from 44 to $\sim 2000$. The hard X-ray band is excellent for finding AGN \citep[e.g.,][]{Koss11}, hot clusters of galaxies \citep[e.g.,][]{Wik12}, cataclysmic variables (CVs), blazars \citep[e.g.,][]{Paliya17} and high mass X-ray binaries in the galactic plane due to the penetrating nature of the X-ray emission. 

Amongst many results, the BAT survey provides the fundamental data for the BAT AGN Spectroscopic Survey (BASS) \citep{Koss17}, and the BASS survey has obtained broad spectral coverage of the AGN population across the electromagnetic spectrum from radio to gamma-rays. The BAT 105-month survey catalog has substantially changed our understanding of the nature of the AGN population from the number of mergers, the low redshift luminosity function \citep{Ananna22a, Ananna22b}, the nature of the host galaxies, the connection to cold gas and star formation \citep{Koss21}. Other work has measured the evolution of blazars \citep{Marcotulli22}, the local supermassive black hole population \citep{Powell22}, and the local number density per solar mass of CVs \citep{Suleimanov22}. 

In addition to the source detection, the BAT data over the past $\sim$20 years have provided long term light curves and spectrum monitering of X-ray binaries\citep[e.g.,][]{Corbet24} and AGNs \citep{Papadakis24, Hinkle21}.

Many of the advances made possible with the BAT survey have relied on ever larger samples of objects, to probe higher redshifts, lower luminosities, finding unusual objects and providing a ``finding chart'' for studies with other X-ray telescopes, such as the the Nuclear Spectroscopic Telescope Array (NuSTAR), the X-Ray Imaging and Spectroscopy Mission (XRISM), Chandra and the X-ray Multi-Mirror Mission (XMM-Newton).

The main updates in this catalog include a new gain-offset calibration file that provides up-to-date calibration, and a new survey instrumental response that is calibrated against the updated Crab spectrum from this analysis. Moreover, we perform additional analysis on the newly emerged systematic noise, which affects our expected sensitivity and source detection.

In addition to our work, the Palermo group has performed independent analysis of the BAT survey data with different data-analysis pipelines, and has published results for the 39-month and 54-month data sets \citep{Segreto_fist_Palermo_a,Cusumano_first_Palermo_b,Cusumano_54m_Palermo}. The Palermo team continues to report results for the 100-month and the 150-month Palermo catalogs on the Palermo BAT catalog website\footnote{\url{https://science.clemson.edu/ctagn/bat-150-month-catalog/}} \citep{Cusumano_100m_Palermo}. 
 
The {\it INTEGRAL}/IBIS team has also recently published the 17-yr hard X-ray all-sky survey catalog \citep{Krivonos22_INTEGRAL}. The {\it INTEGRAL} catalog includes 929 sources and reaches a flux limit of $2 \times 10^{-12} \, \rm erg \, s^{-1} \, cm^{-2}$ for non-blazar AGNs \citep{Krivonos22_INTEGRAL}. 

The paper is organized as follow: Sect.~\ref{sect:data_analysis} describes the data analysis procedure, including updates of the gain-off set calibration and the survey instrumental response. Sect.~\ref{sect:result_instrument} summarizes the characteristics of the 157-month survey data, including the overall exposure and sensitivity, and discussion of how the newly emerged systematic noise affects the detection. Sect.~\ref{sect:result_source} presents the catalog results of the source classification, light curves and spectra. Sect.~\ref{sect:catalog_cross_check} presents results from cross-checking between the 157-month survey catalog and other recent X-ray survey catalogs, including the 150-month Palermo catalog$^{1}$, the 17-yr {\it INTEGRAL} catalog \citep{Krivonos22_INTEGRAL}, and two recent X-ray catalogs in the softer X-ray bands, the $SRG$/ART-XC and $SRG$/eROSITA catalogs\footnote{Upon the completion of this paper, a newer ART-XC catalog has been published recently in July, 2024 \citep{Sazonov24_ART_XC}, which includes comparison with the 105-month survey catalog.} \citep{Pavlinsky21_ARTXC, Merloni24_eROSITA24}.

\section{Data analysis}
\label{sect:data_analysis}


The data reduction and analysis follow the same procedure adopted in previous BAT survey catalogs \citep[e.g.,][]{Tueller10, Baumgartner13}. The analysis uses the BAT survey data, which are collected into the Detector Plane Histograms (DPHs) that are binned in $\sim 300$ s time interval and in 80 energy channels. For the analysis in this catalog, we binned the DPH data into the standard eight energy bands that has been used in all BAT survey data analysis. We choose to adopt the standard procedure instead of using the entire 80 energy channels so that the new data analysis is compatible with procedures in previous BAT survey catalogs, and allows us to remove the detector pattern noise.

We process data from February 2007 to December 2017. In addition to processing data after the 105-month survey catalog \citep{Oh18}, which ends in August 2013, we reprocess some data that were covered by the BAT 105 month survey catalog to include an newly available gain-offset calculation and pattern noise fits (see the pattern noise paragraph below and Sect.~\ref{sect:gain} for more details). All the analyses use HEASoft tools version 6.23, and the most recent BAT calibration database (updated on Oct. 3, 2017). We highlight the main steps here. 

(1) {\it Processing data with batsurvey}: We use the standard BAT pipeline for survey analysis, {\it batsurvey}, to create snapshot images and source detection in eight energy bands (14-20, 20-24, 24-35, 35-50, 50-75, 75-100, 100-150, and 150-195 keV). We compiled an input catalog with sources from (a) the original catalog used in the analysis for previous catalog, (b) new sources found in the 105-month survey catalog, and (c) sources from the BAT transient monitor \citep{Krimm13}. All the newly-added sources with significant flares are marked as ``always clean'' so that the BAT detection algorithm ({\it batcelldetect}) will automatically perform background cleaning for potential contamination from these bright sources. The flares are found either by visual inspection of the light curves, or by searches through ATEL notices. 

(2) {\it Creating pattern noise maps}: As mentioned in the BAT 22-month survey catalog \citep{Tueller10}, pattern noise arises from non-uniform detector properties and contributes to additional systematic noise. The pattern noise becomes noticeable in images with exposure time $\gtrsim$ a day and needs to be subtracted. We adopted the original script used by \citet{Tueller10} to create the pattern noise maps, which are FITS-format image files that contain the value of pattern noise of each detector pixel.
Specifically, the script calculates the average count rate in each detector pixel, and fit a polynomial function to the daily count rate for each pixel. The resulting fitted values are then written in output FITS files to be used in the next step.

(3) {\it Rerunning the survey process to incorporate pattern noise maps}: Once the pattern noise maps are constructed, we run {\it batsurvey} again with these maps, so the pipeline can subtract the effect of pattern noise.

(4) {\it Creating mosaic images}: We create mosaic images from the snapshot images created by the second {\it batsurvey} process. We follow the same mosaicking procedure as those adopted in previous BAT survey catalogs \citep[see detailed description in][]{Tueller10}. Specifically, the process uses six template images across the entire sky (image c0 to c5) in Galactic coordinates. Images c0 to c3 lie along the Galactic plane, and image c1 covers the Galactic Center. Images c4 and c5 cover the top and bottom Galactic Pole, respectively. The mosaic process adds data from individual snapshot image onto these template images, with error propagation and corrections for off-axis effects. We create mosaic images in eight energy bands for each month to produce the monthly light curves. We also generate mosaic images in eight energy bands for the entire 157 months to generate the 157-month spectra for each source in this catalog. 

In addition, following the same procedure adopted in the 70-month and 105-month survey catalogs, we create a Crab-weighted 157-month mosaic image, which gives a weight for each image in the eight energy band before mosaicking. The Crab-weighted image is aimed to increase detection sensitivity for sources with Crab-like spectra. The final Crab-weighted image contain Crab-weighted count $F_{cw}$ calculated by the following equation
\begin{equation}
F_{cw} = \sum^8_{i=1} W_i F_i
\end{equation}
where $F_i$ is the count rate in each of the eight-band image $i$, and $W_i$ is the Crab weight in each energy band, which are [27.000, 35.260, 22.700, 29.444, 21.272, 16.062, 8.449, 2.630] for the eight bands, respectively. A detailed description of how these weights are determined can be found in the 70-month survey catalog \citep{Baumgartner13}. After multiplying by the Crab weight, the Crab-weighted count is expressed in unit of Crab, where $1 \ {\rm Crab} = 2.386 \times 10^{-8} \, \rm{erg \ s^{-1} \ cm^{-2}}$ in the $14-195$ keV band when using the benchmark Crab spectrum 
\begin{equation}
\label{eq:Crab}
F(E) = 10.17 \, E^{-2.15} \ \left(\frac{\rm photon}{\rm cm^2 \, s \, keV}\right).
\end{equation}
This benchmark Crab spectral model is adopted in all BAT survey catalogs.

(5) {\it Source detection (blind search)}: We run the source detection pipeline ({\it batcelldetect}) on the Crab-weighted 157-month mosaic image. The source detection uses a source radius of 15 pixels and background radius of 100 pixels. Detected sources are defined as those with signal-to-noise ratio above $4.8\sigma$, which is the same detection criteria adopted by previous BAT survey catalogs. In this catalog, the detection threshold of $4.8\sigma$ corresponds to $\sim 1$ false detections in our entire sample (see Sect.~\ref{sect:det_sigma} for detailed discussion). There are 205 new sources detected in the 157-month mosaic images and 51 sources that are in our input catalog but are detected by BAT for the first time. 

(6) {\it Position fitting}: We use {\it batcelldetect} again to obtain a refined position for each newly detected source. Specifically, we run {\it batcelldetect} with an updated input catalog that contains these newly detected sources and their initial positions found in the previous step, and set the option ``posfit=YES'' to have the pipeline perform position fitting within a radius of 0.2 degrees (12 arcmin) from the initial position. The radius is set to 0.2 degrees because this is the general BAT position uncertainty for weak sources. 

The Swift name assigned to each of the newly detected sources is based on RA and DEC from the BAT position found in this step. We decide to use the BAT position instead of the counterpart position for assigning the Swift names, because the counterpart position may change if new studies associate the BAT source to a different counterpart. For all sources in previous BAT catalogs, we adopt the same Swift name, even if the BAT position may have changed slightly. 

(7) {\it Flux fitting}: The pipeline {\it batcelldetect} is then used again to obtain the source count rate at each refined position. We create an eight-band spectrum for each source using the count rates found in the eight-band mosaic images, and fit a simple power-law model to the spectrum. Similar to previous catalogs, we adopt a special BAT response function created for mosaic BAT survey product. More specifically, we create a response function calibrated against the Crab nebula. In other words, when using this response function, the eight-band spectrum of the Crab nebula will produce the benchmark Crab spectrum described by Eq.~\ref{eq:Crab}.

(8) {\it Counterpart association}:
For each of the new detections, we perform searches for the potential counterpart and identify the source type (e.g., AGNs, galaxies, pulsars) based on the nature of its counterpart. Our main search uses sources reported in SIMBAD\footnote{\url{http://simbad.u-strasbg.fr/simbad/}} that are within $12$ arcmin ($0.2$ deg) of the BAT position.
If there is an X-ray bright source in SIMBAD that is close to the BAT position, we mark the source as a potential candidate. 

In addition, we cross check the newly-detected sources with the 2XSPS XRT source catalog \citep{2SXPS}. We inspect any 2XSPS sources that are within $12$ arcmin of the BAT position. For each of these sources, we calculate the estimated flux in the BAT survey energy range ($14-195$ keV) extrapolated from the XRT spectral fits. We compare the extrapolated flux with the actual BAT flux calculated from the BAT survey spectrum. We consider the source as a more likely candidate if it satisfies one of the following criteria: (1) the extrapolated flux from the XRT spectrum is comparable or higher than the actual BAT flux, indicating the this source is likely to be detected in BAT, or (2) the XRT flux in the hard band (2 to 10 keV) is higher than $1.0 \times 10^{12} \ \rm erg \ s^{-1} \ cm^{-2}$. An XRT flux below this value would be less likely to come from AGN-like sources. This flux cut is estimated using a typical AGN photon index of 1.7 and assuming an XRT detection of 0.01 count/s (which corresponds to a $\sim 3 \sigma$ detection in a 7 ks exposure time) in a high absorption case with $N(H)=1 \times 10^{24} \ \rm cm^{-2}$. Most of our high-confident counterparts passed both of these criteria. For XRT sources that have a reported flux but do not have a reported XRT spectral index, we use the second criterion to determine the likely counterpart. 

We manually inspect all the candidates found in SIMBAD and in the 2XSPS XRT catalog to determine the most likely counterpart. The relatively large BAT position uncertainty means that often there are more than one source (even when only accounting for X-ray sources) that are within the BAT error region. Thus, We adopt similar flags used in the 70-month survey catalog to indicate the counterpart association strengths. The meaning of these flags is described in the following paragraphs and is also summarized in Table~\ref{tab:ctpt_stren}. All sources found in previous BAT catalogs are marked with flag $=1$, meaning that the counterpart is identified by previous work. Readers are encouraged to check previous catalogs about the association strengths for these sources. Some of these sources in previous catalogs have updated information of their source types and/or redshifts based on new observations and publications. We have included the updated counterpart information in this catalog.

\begin{center}
\begin{table*}
   \caption{The meaning of each flag indicating the Counterpart association strength.}
       \begin{tabular}{|c|c|c|c|}
       \hline
       Flag & \# in catalog & \% & Meaning \\
       \hline
       \hline
       0 & 114 & 6.03 & Confirmed with X-ray imaging \\
       1 & 1671 & 88.51 & Old association held over from previous catalog \\
       2 & 4 & 0.21 & No good hard X-ray source; soft X-ray source \\
       3 & 2 & 0.11 & No X-ray source; source from another waveband \\
       4 & 38 & 2.01 & Educated guess (unavailable/unchecked X-ray image or crowded region) \\
       5 & 5 & 0.26 & No likely counterpart; BAT source on Galactic plane \\ 
       6 & 13 & 0.69 & No likely counterpart; BAT source off Galactic plane\\
       7 & 41 & 2.17 & No association assigned (unavailable/unchecked X-ray image or crowded region) \\  
       \hline
       \end{tabular}   
       \label{tab:ctpt_stren}
\end{table*}
\end{center}

For all the newly detected sources, we first examine whether the candidates in SIMBAD pass the criteria using the XRT spectral fits. If there is only one source in SIMBAD that passes the criteria, we assign this source as the counterpart and assign a counterpart flag of 0. 

If there are multiple SIMBAD sources that pass the XRT spectral-fit criteria, we take into account of the distances from the BAT position and other information found from literature. For example, we assign the counterpart of SWIFT J1804.1\_3421 to IRXS J180408.9-342058 because (1) it is only 0.48 arcmin from the BAT position, which is significantly closer than the other 4 possible candidates that are at least 5.5 arcmin from the BAT position, and (2) IRXS J180408.9-342058 shows previous X-ray outbursts in 2015 \citep{Negoro15}, which is consistent with the flare seen in the BAT light curve. For sources that we found strong evidence of connection between the BAT source and the associated counterpart, we also assign a counterpart flag $=0$ in the table.

Toward the end of our counterpart matching, two new X-ray catalogs from {\it SRG}/ART-XC and {\it SRG}/eROSITA were released \citep{Pavlinsky21_ARTXC, Merloni24_eROSITA24}. We cross-check the BAT sources with these two catalogs and used the information to better determine the confidence of the counterpart association. For example, we associate several counterparts following the counterpart identification in the ART-XC catalog. We also use the detection in ART-XC and eROSITA to help determine the more likely counterparts.  More discussion of cross-checking between recent catalogs are presented in Sect.~\ref{sect:catalog_cross_check}.

A flag $=2$ indicates that there are X-ray sources within the BAT error regions that are likely to be the counterpart (e.g., a Seyfert galaxy or a quasar), but there are no good X-ray sources found in XRT hard-band observation that pass the XRT spectral-fit criteria. For example. we listed the counterpart for SWIFT J1856.3-4309 to be ESO 281-38, because ESO 281-38 is only 0.9 arcmin away from SWIFT J1856.3-4309, and its XRT flux in 0.2-10 keV is $8.1 \times 10^{-13} \ \rm erg \ s^{-1} \ cm^{-2}$, which is only slightly lower than the flux criteria we adopted ($1.0 \times 10^{-12} \ \rm erg \ s^{-1} \ cm^{-2}$) for determining the XRT counterpart. 

A flag $=3$ indicates that the most likely counterpart appears in a waveband other than X ray. For example, we assigned SWIFT J1831.8-3337 to 2MASX J18311470-3336085, a Seyfert 2 Galaxy reported in SIMBAD based on optical and infrared observations. However, no X-ray sources in the 2SXPS within 12 arcminutes pass the XRT spectral-fit criteria, and thus it remains uncertain whether there is enough X-ray emission from 2MASX J18311470-3336085 to make it detectable in BAT. Therefore, this counterpart has an association-strength flag $=3$. For some cases (e.g., SWIFT J0809.0-1028), there are no X-ray sources found because there are no XRT observations at this location. 

We assign a flag $=4$ in the table for the BAT sources that we are able to make an educated guess of the most likely counterpart from several possible sources, but we do not find enough information to have a confident conclusion. For example, XRT observations may be unavailable at the time of the study, or the only source passes the XRT criteria is very close to the edge of the BAT error circle without enough information to robustly determine its association with the BAT source.
Users are recommended to use caution and perform additional investigation of these sources for their physical nature.

If there are no likely X-ray counterpart found in the XRT data, and no likely sources in other waveband, we assign flag$=5$ if the source is on the Galactic plane, and flag$=6$ if the source is off the Galactic plane. Following the definition in the BAT 70-month survey catalog, we consider sources with $|b| \leq 10$ degree to be on the Galactic plane. Most of the cases assigned with flag $=5$ or $6$ are because there are no XRT detection in previous XRT observations. However, occasiontally there are cases with some XRT detections that do not pass the XRT spectral-fit criteria and do not have any other evidence suggesting their association with the BAT source. 

We assign a flag $=7$ if we cannot determine the most likely candidate in the BAT error circle, either because there are many possible sources in the region, or because there is not enough information at the time of the study to make an educated guess of the most likely counterpart. 

The BAT 105-month survey catalog reclassified all the QSO sources into either one of sub-classes of ``Seyfert'' or ``Beamed AGN'' (including both the BL Lacertae objects, BZB, and Flat Spectrum Radio Quasars, BZQs). For the 21 new QSOs found in this catalog, we follow the same criteria (see also Section 2.3 of \citealp{Marcotulli_2022}) and we determine that all the newly-detected QSOs are FSRQs. Therefore, these QSOs are placed in the ``Beamed AGN'' category.

\subsection{Update of the gain calibration}
\label{sect:gain}

As noted in previous BAT survey papers \citep{Baumgartner13, Oh18}, the BAT energy scale (detector gain) has
experienced gradual changes on $\sim$years timescales.  These changes presumably reflect gradual degradation due to
radiation damage within the CZT detectors, and affects the energy calibration in the BAT data analysis.  

The BAT team has been tracking the gain shift and produces a calibration database to correct the shift. The 105-month survey catalog assumed that the gain solution generated for 2008-2013 could be extended as
unchanged. We have now updated the gain solution beyond 2013 and see there were small changes over that time period, and small changes continue up through 2016. We have reprocessed all studied data starting in 2007 to include
the new gain solution. Details about the updated gain solution can be found in Appendix\ref{sect:gain_appendix}.


\subsection{Update of the survey response file}

The spectra in the BAT survey catalog are created from a mosaic image generated by adding up individual snapshot images over 157 months. 
Therefore, a special instrumental response file is required to perform spectral fitting. The response file is created by renormalizing the BAT response matrix so that the Crab eight-band spectrum from the survey catalog will produce the benchmark spectrum described by Eq.~\ref{eq:Crab}. 

The BAT 70-month and 105-month survey catalogs adopted the same response file. However, this response file no longer produces a good fit for the Crab spectrum with the 157-month data. In other words, the fit using the old response file differs from the benchmark Crab spectrum. Therefore, we update the response file using the Crab spectrum found in the BAT 157-month survey. The response file is created using the original script that produces the response file for the previous BAT catalogs.

\section{Characteristics of the 157-month survey data}
\label{sect:result_instrument}

\subsection{BAT exposure}

\begin{figure*}[!ht]
\begin{center}
\includegraphics[width=1.0\textwidth]{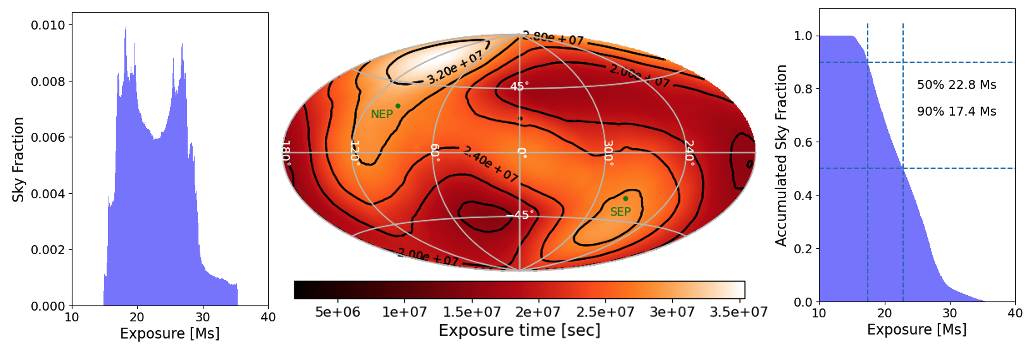}
\end{center}
\caption{
Exposure in the 157-month survey catalog. {\it Left Panel}: the distribution of sky fraction vs exposure time. {\it Middle Panel}: The all-sky map of exposure time in Galactic coordinate. The NEP and SEP mark the North and South Ecliptic Pole, respectively. {\it Right Panel}: The accumulated sky fraction vs. exposure time (i.e., sky fraction that has exposure time greater than the specific number). 
}
\label{fig:BAT_exposure}
\end{figure*}

Figure \ref{fig:BAT_exposure} shows the exposure time in the 157-month survey catalog.  The left panel shows the distribution of the sky fraction as a function of exposure time. The middle panel shows the all-sky map of exposure time in Galactic coordinates. The right panel presents the accumulated sky fraction as a function of exposure time. The exposure time ranges from $\sim 15$ Ms to $\sim 35$ Ms, and $\sim 50\%$ of the sky achieves an exposure time of $\sim 22.8$ Ms. 

 As shown in the middle plot, the North and South Ecliptic Pole have larger exposure times due to the Sun and Moon observing constraints. In other words, BAT spends less time on the Ecliptic Plane as {\it Swift} cannot point too close toward the Sun and the Moon. The North Ecliptic Pole has slightly larger exposure time than the South Ecliptic Pole, which is likely due to another observing constraints around the South Atlantic Anomaly (SAA).

\subsection{BAT sensitivity}
\label{sect:BAT_sensitivity}

\begin{figure}[!h]
\begin{center}
\includegraphics[width=0.5\textwidth]{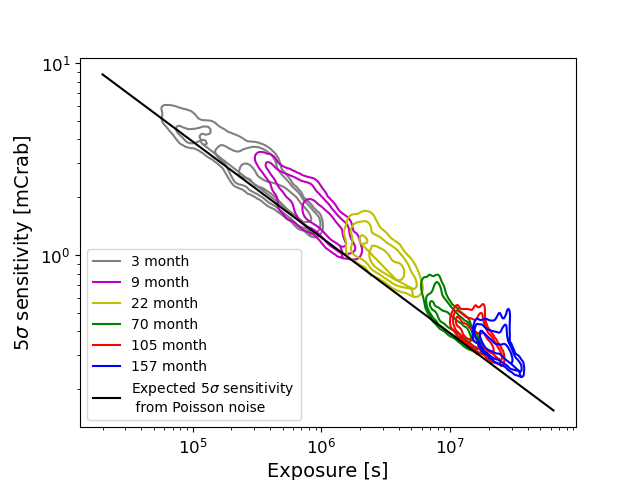}
\end{center}
\caption{
The $5\sigma$ sensitivity for each pixel in the BAT mosaic images as a function of effective exposure time. Different colors represent sensitivities from different BAT survey catalogs, and the contour lines are values that enclose 1\%, 10\%, and 50\% of pixels. The black line shows the expected $5\sigma$ sensitivity from Poisson noise.
}
\label{fig:BAT_sensitivity}
\end{figure}

\begin{figure}[!h]
\begin{center}
\includegraphics[width=0.4\textwidth]{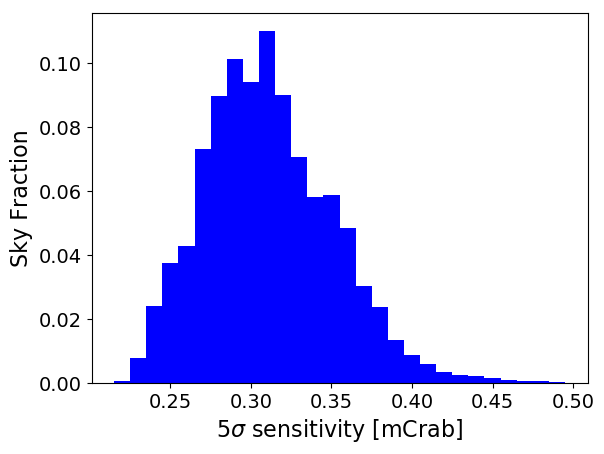}
\includegraphics[width=0.4\textwidth]{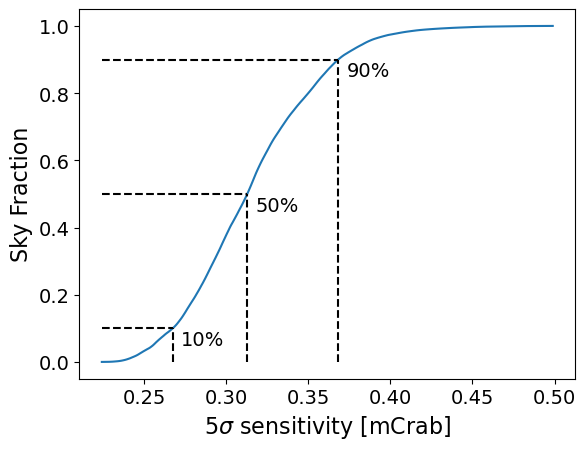}
\end{center}
\caption{
{\it Upper panel}: Sky coverage histogram for the $5\sigma$ sensitivity.
{\it Bottom panel}: Accumulated sky fraction vs. $5\sigma$ sensitivity.
}
\label{fig:BAT_sensitivity_skyfrac}
\end{figure}

Figure \ref{fig:BAT_sensitivity} summarizes the $5\sigma$ sensitivity for each pixel in the BAT mosaic image as a function of exposure time. Different colors represent the sensitivities from different BAT survey catalogs \citep{Markwardt05, Tueller08, Tueller10, Baumgartner13, Oh18}. The contour lines for each catalog show regions that enclose 1\%, 10\%, and 50\% of pixels.  The $5\sigma$ sensitivity is calculated from the noise level estimated using the BAT HEASoft tool, {\it batcelldetect}, which calculates the local background root-mean-square values in the mosaic image. Following previous catalogs, we use the Crab-weighted mosaic image for the noise estimation, and we exclude a circular region with a radius of 15 pixels around each always-clean source, as the cleaning process produces slightly different background level around the source location.

The black line in Fig.~\ref{fig:BAT_sensitivity} is the expected BAT sensitivity as a function of exposure time based on the following equation estimated in the BAT 70-month catalog \citep{Baumgartner13},
\begin{equation}
\label{eq:sensitivity}
    f_{5\sigma} = \rm 1.18 mCrab \Big( \frac{T}{\rm 1 Ms} \Big)^{(-1/2)},
\end{equation}
where $f_{5\sigma}$ is the flux limit for a $5\sigma$ detection and $T$ is the ``effective on-axis exposure time''. That is, $T = pT_0$, where $p$ and $T_0$ are the partial coding fraction and the actual exposure time at a specific location. This equation is calculated from the expected $5\sigma$ Poisson noise level derived in \citet{Skinner08}, 
\begin{equation}
\label{eq:skinner08}
5\sigma_{\rm Poisson} = 5 \sqrt{\frac{2b}{\alpha N_{\rm det} T}},
\end{equation}
where $b=0.246 \ \rm cts \ s^{-1} \ detector^{-1}$ is the per-detector count rate that includes background and point sources in the field of view, $\alpha = 0.27$ is a coefficient related to the mask pattern and detector pixel size, and $N_{\rm det} = 22520$ is the number of enabled detectors. The values are determined in the BAT 70-month survey catalog \citep{Baumgartner13}. 
The $5\sigma$ BAT sensitivity is derived from the perspective of pure Poisson counting statistics. Therefore, this equation (and thus the black line) represents only the statistical uncertainty, but does not include systematic uncertainty. 

As seen in the figure, the BAT sensitivity has followed closely with this expected sensitivity, indicating that statistical uncertainty dominates over systematic uncertainty. However, 157-month mosaic image shows that the BAT sensitivity starts to diverge from the expected statistical noise, implying the emergence of additional noise. 
One possibility is that BAT has additional statistical noise in recent years, because there are fewer enabled detectors as more and more detectors became noisy and are permanently turned off. The normalization factor in Eq.~\ref{eq:sensitivity}, 1.18 mCrab, is calculated from Eq.~\ref{eq:skinner08}. A smaller number of $N_{\rm det}$ will correspond to a larger normalization factor, and thus decrease the gap between the the black line and the 157-month contours. $N_{\rm det}$ has decreased from 29413 in 2005 to 18504 in 2017, with an average enabled detector $N_{\rm det}=20088$ throughout this period. If we use this updated $N_{\rm det}$, the normalization factor in Eq.~\ref{eq:sensitivity} changes from 1.18 mCrab to 1.23 mCrab, which can explain some of the extra noise level, but not all.
Therefore, part of the additional noise is likely from systematic noise. 

At least some of the systematic noise could be introduced by the {\it batclean} process to clean (reduce) the contamination from bright sources. This is because the {\it batclean} process results in a different noise level around the bright sources. However, {\it batclean} has been adopted in many previous BAT survey catalogs, which did not show any significant systematic noise. Therefore, we suspect that the main cause of the additional systematic noise is likely to be intrinsically instrumental. As the {\it Swift} mission lasts longer and the total exposure time in the mosaic image increases, we expect to see the intrinsic instrumental systematic noise at some point, and the BAT sensitivity will not improve indefinitely with the increase of exposure time. It is possible that we have started to see this instrumental systematic noise in the 157 month processing.
This additional noise results in $\sim 50\%$ fewer new detections than we expected (see more discussion in Sect.~\ref{sect:new_detection}).

The overall $5\sigma$ sensitivity for the 157-month survey catalog is summarized in Figure \ref{fig:BAT_sensitivity_skyfrac}. The upper panel shows the distribution of the sky fraction as a function of the $5\sigma$ sensitivity. The bottom panel shows the accumulated sky fraction as a function of the $5\sigma$ sensitivity. The $5\sigma$ sensitivities for $10\%$, $50\%$, and $90\%$ sky coverage are 0.27 mCrab ($6.44 \times 10^{-12} \ \rm erg \ s^{-1} \ cm^{-2}$), 0.31 mCrab ($7.40 \times 10^{-12} \ \rm erg \ s^{-1} \ cm^{-2}$), and 0.37 mCrab ($8.83 \times 10^{-12} \ \rm erg \ s^{-1} \ cm^{-2}$), respectively. 

\subsection{Detection significance}
\label{sect:det_sigma}

The standard detection threshold of $4.8 \sigma$ was first set in the 9-month BAT survey catalog \citep{Tueller08} and the last detailed study of the false-detection rate for this threshold was done in the 22-month BAT survey catalog \citep{Tueller10}. Since over a decade has passed, we decided to re-investigate the false-positive rate associated with the $4.8 \sigma$ threshold.

\begin{figure}[!h]
\begin{center}
\includegraphics[width=0.5\textwidth]{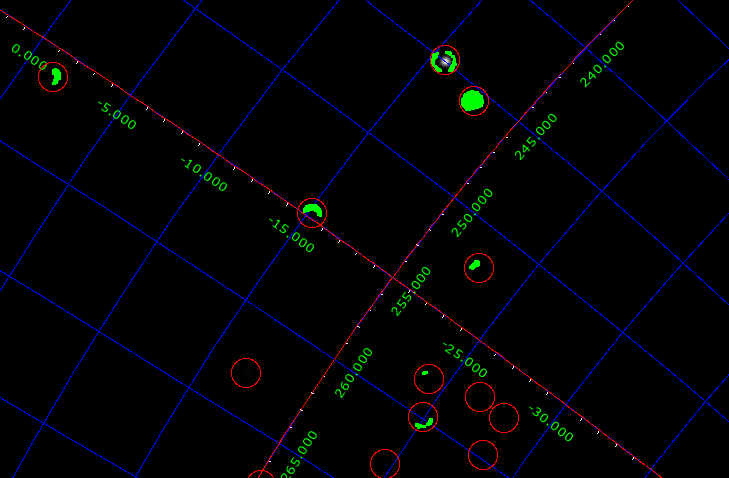}
\end{center}
\caption{
Example of some artifacts from batclean (clusters of negative SNRs) when attempting to remove potential contamination of bright sources in the image. 
The red circles mark a region of 15 pixels ($\sim 40$ arcmin) around the always-clean sources (i.e., bright sources that will always be cleaned by the analysis), and the green dots indicate locations where SNR $< -4.8\sigma$. The RA and Dec coordinate grid is presented to show the scale of this map.
}
\label{fig:c1_example}
\end{figure}

Following the same procedure adopted in the 22-month BAT survey catalog, we calculate the signal-to-noise ratio (SNR) from the mosaic sky map, which includes both statistical and systematic noise.
We exclude any pixels that are within a radius of 15 pixels ($\sim 40$ arcmin) from any sources that are marked as always clean. As mentioned in Sect.~\ref{sect:data_analysis}, these are sources that are either bright or have known bright flares, and thus will always be applied to the {\it batclean} pipeline to remove their potential contamination to other sources. However, the {\it batclean} process can introduce incorrect background variation around the cleaned sources, and are known to generate some artifacts such as an excess of negative SNRs located around the cleaned sources\footnote{More specifically, the batclean procedure uses ray tracing to estimate the locations of photons from the bright source on the detector plane, and subtracts those photons on the detector plane before creating a mask-weighted image. However, it is known that sometime batclean can produce residuals in the coded-masked images, which will result in some negative SNRs positions clustered around the cleaned source. 
}. An example can be seen in Fig.~\ref{fig:c1_example}, where the red circles mark a radius of 15 pixels around the always-clean sources, and the green dots mark the pixels with $\rm{SNR}< -4.8 \sigma$. Almost all the green dots lie within red circles, as they are likely from the artifact of {\it batclean} and not from the Gaussian fluctuation. Therefore, these regions are not included in our false-detection rate studies (nor are they used in our source detections).


\begin{figure}[!h]
\begin{center}
\includegraphics[width=0.45\textwidth]{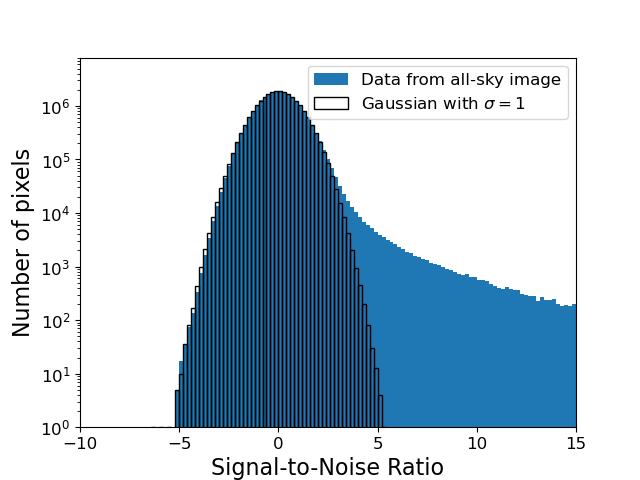}
\end{center}
\caption{
Histogram of the SNR from pixels in the 157-month mosaic significance map. The black shaded region is a histogram made by randomly generated numbers that follows a $1-\sigma$ Gaussian distribution. 
}
\label{fig:BAT_snr_hist}
\end{figure}

The blue bars in Fig.~\ref{fig:BAT_snr_hist} shows the resulting SNR histogram measured from the 157-month mosaicked sky image. For comparison, the black shaded region shows a $1-\sigma$ Gaussian distribution, 
normalized to the peak of the distribution from actual data (the blue bars). As shown in the plot, the negative SNR side in the histogram from actual data follows very closely to the expected $1-\sigma$ Gaussian distribution, indicating that the false-detection rate should still generally follows a Gaussian distribution. Therefore, in the positive SNR side, the excess counts in the histogram compares to the Gaussian distribution are likely from real detections. 


A $4.8\sigma$ detection for a perfectly Gaussian distribution would correspond to a false-detection rate of $7.93 \times 10^{-7}$, which would have translated to less than one false detection in the BAT all-sky image of $\sim 10^6$ physical pixels\footnote{The BAT mosaic images consist of six images covering six facets in the sky. Each facet image consists of $1998 \times 1998$ pixels. However, there are $\sim 20\%$ overlap between each facet image. Therefore, the number of image pixels for entire sky is $\sim 1998 \times 1998 \times 0.8 = 1.9 \times 10^{7}$. In our search, sources within BAT's angular resolution of $12$ arcmin (4.26 image pixel) are considered the same source. Therefore, the number of physical pixels in our search is $\sim 1.9 \times 10^7/(4.26 \times 4.26) = 1.1 \times 10^6$.}, as noted in \citet{Tueller10}. The actual fraction of pixels with $\rm{SNR} < -4.8 \sigma$ in our image is 
$9.19 \times 10^{-7}$
(since the pixels with $\rm{SNR} > 4.8 \sigma$ includes real detections, we use the pixels with $\rm{SNR} < -4.8 \sigma$ for the false-detection rate estimation). This fraction is only slightly higher than the fraction from a perfect Gaussian, and corresponds to $\sim 1$ false detection in our sample.


\subsection{Newly detected sources}
\label{sect:new_detection}


There are 256 newly-detected sources in the BAT 157-month survey catalog, of which 51 sources are previously known sources with potential hard X-ray emission, and are in our input catalog when performing the source detection. The total number of newly-detected source is $\sim 50\%$ less than what was originally expected based on the following order-of-magnitude estimation, which assumes the BAT sensitivity has not reached the very tail end of the source luminosity distribution. In other words, we assume BAT has not detected all the dim sources yet. 

BAT is a photon-counting instrument, the signal from a source (i.e., photon counts $N_{\rm ph}$ from a source) is proportional to the exposure time $T$, and the noise is proportional to $\sqrt{N_{\rm ph}}$. Thus, the signal-to-noise ratio is proportional to $N_{\rm ph}/\sqrt{N_{\rm ph}} = \sqrt{N_{\rm ph}}$ and is therefore proportional to $\sqrt{T}$. As a result, the flux limit $f_{\rm limit}$ for a detectable source decreases as $1/\sqrt{T}$, i.e., $f_{\rm limit} \propto T^{-{1/2}}$. 

Because the flux decreases with the square of the luminosity distance of a source, and thus $f_{\rm limit} \propto \, {\rm D}_{\rm max}^{-2}$, where ${\rm D}_{\rm max}$ is the maximum distance for a source to be detectable by BAT. Therefore, ${\rm D}_{\rm max} \propto T^{1/4}$. Assuming the number of sources is proportional to the volume of the space (i.e., we assume a flat luminosity distribution for simplicity), the expected number of detections would be $N \propto V_{\rm max} \propto D^{3}_{\rm max} \propto T^{3/4}$. The 105-month survey catalog detects 1632 sources, and thus the expected number of detections in the 157-month survey catalog is $1632 \times (157/105)^{3/4} = 2206$. This gives us $2206-1632 = 574$ expected newly detected source in this catalog, about $2.2$ times higher than what is actually detected.

We perform extensive investigation of the reasons for this unexpected decrease in the number of detection, and we conclude that the main cause is the additional systematic noise that starts to appear in the 157 month mosaic image.

\begin{figure}[!h]
\begin{center}
\includegraphics[width=0.46\textwidth]{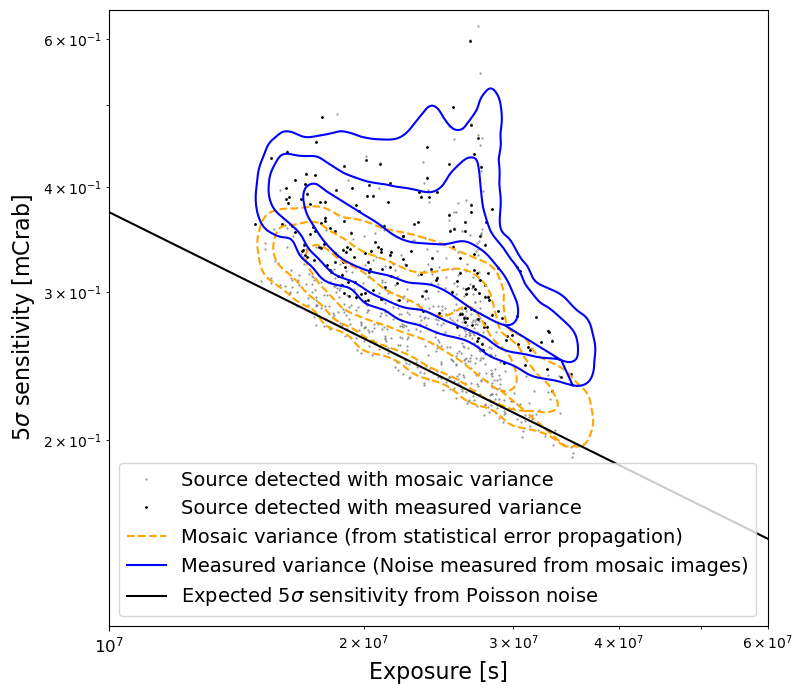}
\end{center}
\caption{
The $5\sigma$ sensitivity contours from the BAT mosaic images as a function of effective exposure time. The dark blue contours (solid line) show the $5\sigma$ sensitivity measured from the actual BAT mosaic image, and thus include statistical and systematic noise. The orange contours (dashed line) show the $5\sigma$ sensitivity measured from the mosaic variance map, which is estimated from error propagation based on the noise level in each snapshot image, and thus does not include correlated noise. The gray dots are detected sources using the mosaic variance map. The black dots are actual detected sources in the 157-month survey catalog. 
}
\label{fig:BAT_sensitivity_compare}
\end{figure}

As seen in Fig.~\ref{fig:BAT_sensitivity}, the improvement of the BAT sensitivity in the 157-month survey catalog is less than expected from pure statistical noise. 
To further investigate the origin of this additional noise and how it impact the number of detection, we plot the $5\sigma$ BAT sensitivity estimated from the mosaic variance map in Fig.~\ref{fig:BAT_sensitivity_compare} (orange-dashed contour). As mentioned earlier, the mosaic variance map is created by adding up the variance measured in individual snapshot images by standard error propagation. Therefore, the noise level from the mosaic variance map does not include any correlated noise. As seen in Fig.~\ref{fig:BAT_sensitivity_compare}, the $5\sigma$ sensitivity from the mosaic variance map is back on track with the estimation based on statistical noise (the black line; Eq.~\ref{eq:sensitivity}). The blue contour in Fig.~\ref{fig:BAT_sensitivity_compare} is the same as the blue contour shown in Fig.~\ref{fig:BAT_sensitivity}, which is based on the variance measured directly from the 157-month mosaic image. 

The source detection pipeline {\it batcelldetect} usually calculates the SNR based on the local variance measured in the same mosaic image. However, the pipeline has an option to use different variation values by providing an input variation map.
When we run {\it batcelldetect} using the mosaic variation map as the input variation, the pipeline detects $743$ new sources, which is more comparable (1.3 times) with the expected number from the order-of-magnitude estimation described above that only accounts for statistical noise. These detections are plotted as grey dots in Fig.~\ref{fig:BAT_sensitivity_compare}. Both these grey dots and the $5\sigma$ sensitivity from the mosaic variation map follows the statistical expectation, implying that the additional noise comes from correlated noise (i.e. some kind of systematic noise). Note that the increase in number of detections when using the mosaic variation map is because we ignore the systematic noise, and thus these additional detections are most likely to be false detections. 
For comparison, the actual new detections in the 157 months are plotted as black dots.


\subsection{Source positions and Uncertainties}
\begin{figure}[!h]
\begin{center}
\includegraphics[width=0.5\textwidth]{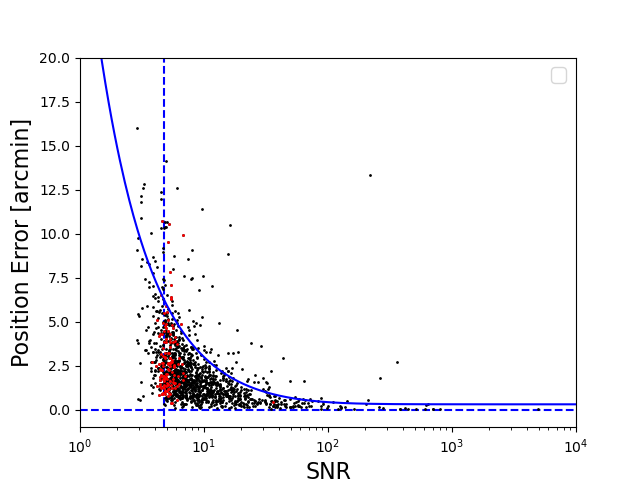}
\end{center}
\caption{
The BAT position error (i.e., the angular separation between the BAT position and the counterpart position) as a function of SNR of the detected source. The solid blue line shows the empirical function that encloses $90\%$ of the sources. The red dots are the newly-detected sources in the 157-month survey catalog. The vertical dashed line shows the detection threshold of $4.8 \sigma$. Some red sources fall slightly below this line because the SNR reported here are from the final flux fit, and thus the value may vary slightly from the original blind search.
}
\label{fig:pos_err}
\end{figure}

Similar to previous BAT catalogs, we examine the accuracy of the BAT position by plotting the angular separation between the BAT position and the counterpart position against the SNR of the BAT detection. The result is shown in Fig.~\ref{fig:pos_err}. The red dots are the newly-detected sources in the BAT 157-month survey catalog. The vertical blue dashed line plots the $4.8 \sigma$ detection threshold. Note that some of the newly detected sources have SNR that are slightly less than $4.8 \sigma$, because the value plotted here are from the final flux fit (see description in Sect.~\ref{sect:data_analysis}), and thus some SNR values may change slightly from the initial blind search. We choose to use the position and SNR from the final flux fit, since these positions are obtained after the position-fitting step (following the initial blind search) and thus should be more accurate. In addition, there are some previously detected sources that have SNR values much lower than $4.8 \sigma$, these are likely to be variable sources (e.g., they may have flares during the time of previous catalogs) and thus the SNR decreases in the 157-month survey catalog.

The solid blue line in the figure is an empirical function described by the following equation,
\begin{equation}
{\rm BAT \ error \ radius \ [arcmin]} = \sqrt{\left(\frac{29.7}{{\rm SNR}}\right)^2+(0.3)^2}.
\end{equation}
Specifically, we adopt the same functional form as those in previous BAT catalogs. We include a systematic error of 0.3 in the equation to incorporate the position errors for those sources with high SNR. This systematic error is slightly larger than the number (0.1) reported in the 105-month survey catalog, but it is identical to the one reported in the 70-month survey catalog. Once the systematic error is set, we modify the only variable in the equation until the function capture $90\%$ of all the sources. We obtain a value of 29.7 for this variable, which is very similar to those reported in the 105-month catalog (30.5) and in the 70-month catalog (28).

\section{The {\textit Swift}/BAT 157-month survey catalog}
\label{sect:result_source}

\subsection{Summary of source types}
\begin{figure*}[!h]
\begin{center}
\includegraphics[width=0.8\textwidth]{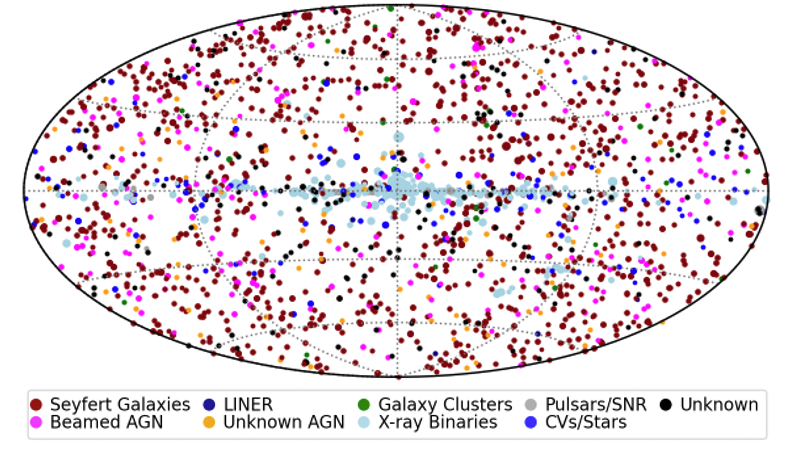}
\end{center}
\caption{
An all-sky map of the 1888 sources in the BAT 157-month survey catalog. There are 256 newly-detected sources. A larger marker size represents sources with higher flux.
}
\label{fig:BAT_allsky_map}
\end{figure*}

\begin{figure*}[!h]
\begin{center}
\includegraphics[width=0.52\textwidth]{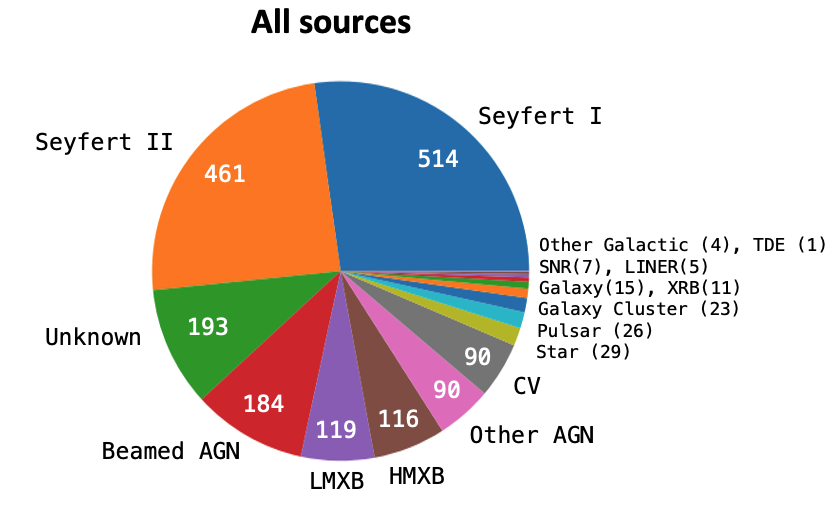}
\includegraphics[width=0.47\textwidth]{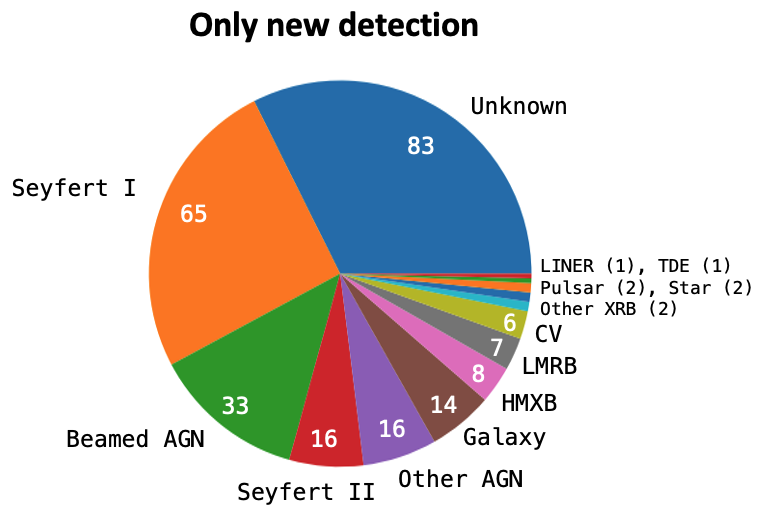}
\end{center}
\caption{
Distributions of source types for all sources in the 157-month survey catalog (left panel), and for newly detected sources (right panel).
}
\label{fig:Pie_chart}
\end{figure*}

We follow similar source type classification adopted in the 70-month survey catalog \citep{Baumgartner13} and the 105-month survey catalog \citep{Oh18}. The 105-month survey catalog presents a more detailed classification than those in the 70-month survey catalog and previous catalogs. For example, the 105-month survey catalog places quasars into either Seyfert or Beamed AGNs, and includes some detailed source types such as the symbiotic stars, molecular clouds, and star clusters. In the 157-month survey catalog, we adopt basic source types from the 70-month survey catalog, but also reclassified quasars into either Seyfert or Beamed AGNs. We keep the descriptions of detailed source types in the 105-month survey catalog, but relabeled them with source flags that are more similar to the 70-month survey catalog. Specifically, symbiotic stars, novae, star clusters are grouped into Class 12 (``Star'').``Galactic Center'', ``molecular cloud'' and ``gamma-ray source'' are marked as Class 1 (``Galactic''). We added one new category of ``Tidal Disruption Event'' (TDE) because an TDE event, SWIFT J164449.3+573451, is detected in the 157-month survey catalog. A summary of the source types we use in this catalog is shown in Table~\ref{tab:source_type}. Figure \ref{fig:BAT_allsky_map} presents the all-sky map of sources in this catalog, and Fig.~\ref{fig:Pie_chart} shows the pie chart of different source categories.

\begin{table*}
   \begin{center}
   \caption{Source types for the counterparts in the BAT 157-month survey catalog}
       \begin{tabular}{|c|c|c|c|}
       \hline
       Class & Source Type & \# in catalog & \% \\
       \hline
       \hline
        0 & Unknown & 193 & 10.22 \\
        1 & Galactic & 4 & 0.21 \\
        2 & Galaxy & 15 & 0.79 \\
        3 & Galaxy Cluster & 23 & 1.22 \\
        4 & Seyfert I & 514 & 27.22 \\
        5 & Seyfert II & 461 & 24.42 \\
        6 & Other AGN & 90 & 4.77 \\
        7 & Beamed AGN (Blazar/FSRQ) & 184 & 9.75 \\
        8 & LINER & 5 & 0.26 \\
        9 & Cataclysmic Variable Star (CV) & 90 & 4.77 \\
        10 & Pulsar & 26 & 1.38 \\
        11 & Supernova Remnant (SNR) & 7 & 0.37 \\
        12 & Star & 29 & 1.54 \\
        13 & High Mass X-ray Binary (HMXB) & 116 & 6.14 \\
        14 & Low Mass X-ray Binary (LMXB) & 119 & 6.30 \\
        15 & Other X-ray Binary (XRB) & 11 & 0.58 \\
        16 & Tidal Disruption Event & 1 & 0.05 \\
       \hline
       \end{tabular}   
       \label{tab:source_type}
    \end{center}
\end{table*}

\subsection{BAT fluxes and spectra}

For each detected source, we extracted the count rate at the BAT position in the eight-band images and create the eight-band spectra. More specifically, we use the ``CENT\_RATE'' from the output of {\it batcelldetect} 
 (except for sources with potential contamination from nearby sources.) The CENT\_RATE is the background-subtracted source count rate derived from the nearest pixel of the input source position, and is recommended by the BAT team to be used for source count estimation in mosaic images and for faint sources. In comparison, the ``RATE'' reported by {\it batcelldetect} is derived by fitting a point-spread function, which may not represents the true point-spread function in mosaic images and for faint sources. Occasionally, several sources may be very close in position and the ``CENT\_RATE'' may have contribution from all these sources. We examine these sources and use the ``RATE'' values when necessary, as the point-spread-function fitting could better disentangle contamination from nearby sources. Sources with potential contamination are marked as ``Y'' in the Contamination column in the summary table. 

We fit the eight-band spectra with with the {\it pegpwrlw} model (power law with pegged normalization) using Xspec (version 12.11.1). We use the error command in Xspec to find the 90\% confident region for the $14-195$ keV flux and the power-law index.

\begin{figure*}[!h]
\begin{center}
\includegraphics[width=1.0\textwidth]{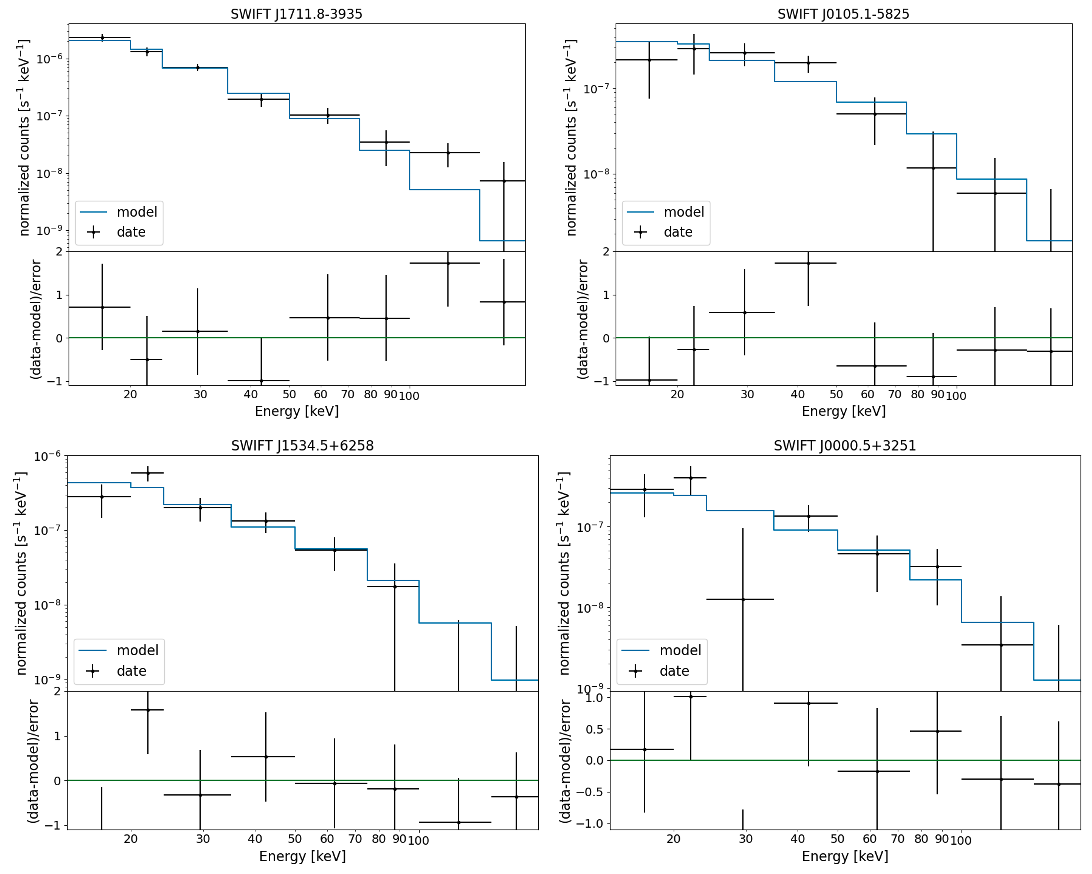}
\end{center}
\caption{
Some examples of the spectra and spectral fits from the 157-month survey catalog. Spectra for all the sources in the catalog can be found in the online journal and on the 157 month web page.
}
\vspace{-5pt}
\label{fig:spectra}
\end{figure*}

Figure \ref{fig:spectra} shows some examples of the spectra and spectral fits. The complete data set can be found in the online journal and on the 157 month web page.

\subsection{Light Curves}

The 157 month survey catalog contains three different types of monthly light curves: (1) the eight-band light curve, (2) the Crab-weighted light curve, and (3) the snapshot light curve.

The eight-band light curves are created by running the BAT source detection algorithm {\it batcelldetect} on the eight-band monthly-mosaic images with input BAT source positions. The count rate reported in this light curve is again the ``CENT\_RATE'' estimated by {\it batcelldetect} (except for sources with potential contamination from nearby sources), and the rate error is the ``BKG\_VAR'' (background variation) calculated by {\it batcelldetect}.

The Crab-weighted light curves are created by running {\it batcelldetect} on the monthly Crab-weighted mosaic images with input BAT source positions. The Crab-weighted mosaic images are created by the eight-band monthly mosaic images with the same Crab-weights described in Sect.~\ref{sect:data_analysis}. 
The $14-195$ keV light curve plots shown in the BAT 157-month survey catalog web page are the Crab-weighted light curves. Thus, the count rate unit of the light curve is marked as ``Crab''. However, please note that this simply means that it is a weighted sum of count rate in each of the eight energy band using the Crab weights, and does not mean the light curve is normalized to the entire Crab light curve. In fact, the Crab light curve itself shows some intrinsic variation around one Crab unit \citep[see, e.g.,][]{Oh18, Wilson-Hodge11}. 

The snapshot light curves are created by running {\it batcelldetect} on the snapshot images with input BAT source positions. The snapshot images are created by {\it batsurvey} using the BAT DPH data. Since these are not mosaic images, the source detection method uses the default source radius of $6$ pixels and background radius of $30$ pixels, which is different than the values of $15$ and $100$ pixels used in the source detection in the mosaic images (see descriptions in Sect.~\ref{sect:data_analysis}).

\begin{figure*}[!h]
\begin{center}
\includegraphics[width=1.0\textwidth]{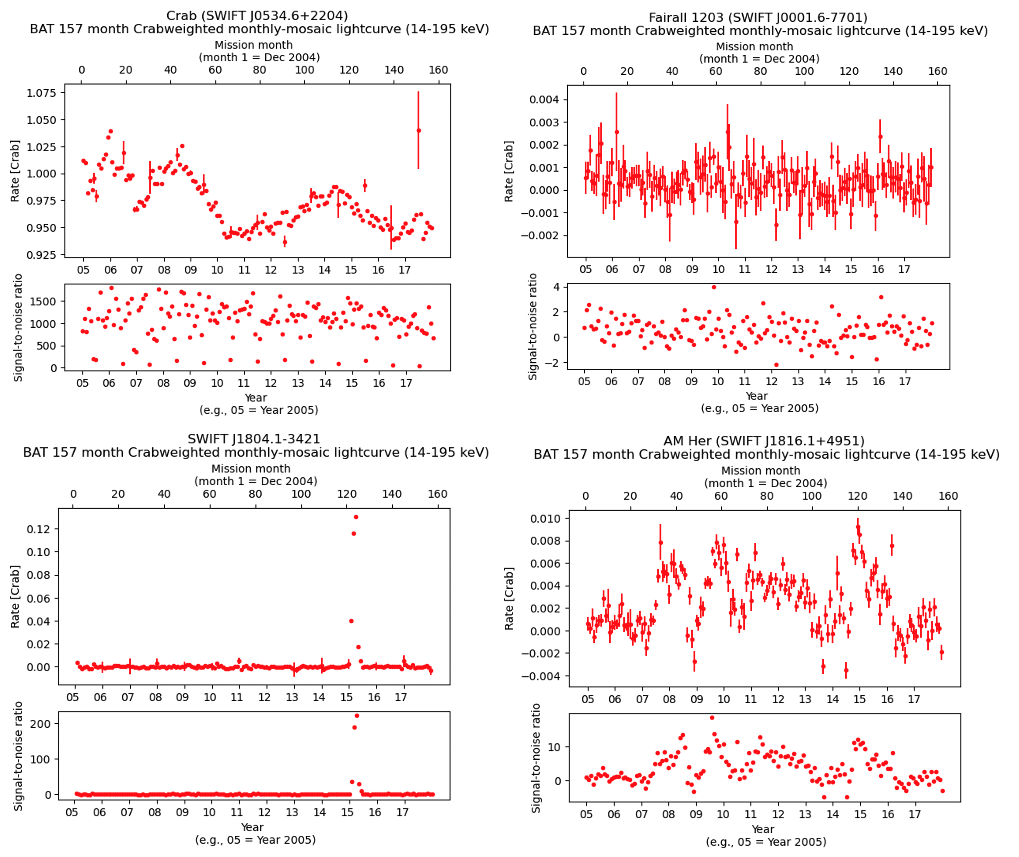}
\end{center}
\caption{
Some examples of the Crab-weighted monthly light curves from the 157-month survey catalog. The complete light curves can be found in the online journal and on the 157 month web page.
}
\label{fig:lightcurves}
\end{figure*}

Figure \ref{fig:lightcurves} shows some examples of the Crab-weighted monthly light curves. The complete light curves data can be found in the online journal and on the 157 month web page.

\section{Comparison with recent X-ray catalogs from other missions}
\label{sect:catalog_cross_check}

\subsection{The 17-yr {\it INTEGRAL}/IBIS catalog}

The {\it INTEGRAL}/IBIS catalog reports a sensitivity that measures AGN down to $\sim 2 \times 10^{-12}  \ \rm erg \ s^{-1} \ cm^{-2}$ \citep{Krivonos22_INTEGRAL}, which is lower than the sensitivity for the BAT 157-month survey catalog for 90\% of the sky coverage ($8.86 \times 10^{-12} \ \rm erg \ s^{-1} \ cm^{-2}$). However, the {\it INTEGRAL} catalog adopts a detection threshold of $4.5 \sigma$, which is also slightly lower than the $4.8 \sigma$ adopted in the BAT catalog. 
As mentioned in Sect.~\ref{sect:intro}, {\it INTEGRAL}/IBIS sky survey is complementary with the BAT sky survey, as {\it INTEGRAL}/IBIS has more observations near the Galactic plane and most of the BAT observations are away from the Galactic plane.   
Out of the 1888 sources, 741 sources are within $12$ arcmin of an {\it INTEGRAL} source. Out of the 256 newly detected sources in the 157-month survey catalog, 30 sources are also reported in the {\it INTEGRAL}/IBIS 17-yr catalog, 9 of these sources are on Galactic plane $|b| \leq 10$ deg).

\subsection{The 150-month Palermo catalog}

As mentioned in Sect.~\ref{sect:intro}, the Palermo BAT survey catalog were carried out by an independent effort using different data analysis pipeline  \citep[BATIMAGER; see more details in][]{Segreto10}, but also adopted a $4.8 \sigma$ detection threshold. The recent Palermo 150-month BAT survey catalog includes 1390 sources that are not on the Galactic plane (i.e., $\rm{|b| > 10 \ \rm deg}$). In comparison, there are 1377 sources in the BAT 157-month survey catalog with $\rm{|b| > 10 \ \rm deg}$, within which 1060 (77\%) sources are within $12$ arcmin from a Palermo source. This overlapping fraction is higher than the $68\%$ found when comparing the 105-month survey catalog with the Palermo 54-month catalog \citep{Oh18, Cusumano_54m_Palermo}.
In addition, the 157-month BAT catalog catalog contains $317$ sources that are not in the Palermo 150-month catalog. On the other hand, the Palermo 150-month catalog includes $341$ sources that are not in the 157-month BAT catalog. The difference may be due to different data processing methods, and different amount of data included in the analysis (157 months vs 150 months),


\subsection{The SRG/ART-XC and SRG/eROSITA all-sky X-ray survey}
The ART-XC and eROSITA are both telescopes onboard the $SRG$ observatory, which was launched in July 13, 2019. The ART-XC telescope covers an energy range of $4-12$ keV, while the eROSITA telescope covers $0.2-5$ keV. Catalogs from the first-year all-sky survey from these two telescopes were released recently \citep{Pavlinsky21_ARTXC,Merloni24_eROSITA24}. In the first-year catalog, ART-XC is able to detect point sources as dim as $\sim 4 \times 10^{-12} \ \rm erg \ s^{-1} \ cm^{-2}$ around the ecliptic plan and $\sim 8 \times 10^{-13} \ \rm erg \ s^{-1} \ cm^{-2}$ around the ecliptic pole \citep{Pavlinsky21_ARTXC}.

Out of 867 sources in the ART-XC catalog, 19 of the newly-detected sources in the BAT 157 month catalog lie within $12$ arcmin of sources in the ART-XC. All of these 19 sources have one-to-one association with the BAT sources, and thus we consider these as likely counterparts to the BAT sources. Specifically, we list the associated ART-XC source as the counterpart for three BAT sources (SWIFT J0716.8-7105, SWIFT J1906.5-0446, and SWIFT J0158.2-8400), since these three sources do not have any other identified counterparts. Moreover, for four BAT sources (SWIFT J0613.4-2903, SWIFT J1331.4-5456, SWIFT J1622.0+5430, and SWIFT J1656.0+2117), we adopt the counterpart identified with the ART-XC sources, because there are no XRT observations available at the time of the study and we cannot identify the most likely counterparts.

Upon the completion of this paper, we noticed that there is an newer ART-XC catalog published recently in July, 2024 \citep{Sazonov24_ART_XC}, and it includes the first five surveys by ART-XC. This paper reports 785 cross matches between the new ART-XC catalog and the BAT 105 month catalog \citep{Oh18}. 

The SRG/eROSITA all-sky survey 
includes three catalogs: (1) the ``main'' catalog that includes all detected X-ray sources in the $0.2-2.3$ keV band, (2) a catalog that includes point-like X-ray sources detected in the $0.2-2.3$ keV band, and (3) a ``hard'' catalog that includes hard X-ray sources detected in the $2.3-5$ keV band \citep{Merloni24_eROSITA24} . Due to the large error region of BAT, there are often many eROSITA sources within the BAT error circle. Therefore, we focus on the results from cross-checking with the ``hard'' catalog. Out of the 1888 sources in the 157-month survey catalog, $620$ sources have eROSITA sources found within 12 arcmin of the BAT source. Within which, $545$ BAT sources match to only one eROSITA source, and $75$ BAT sources have more than one eROSITA sources within the error circle. Because the eROSITA catalog only covers the Western Galactic Hemisphere, this means roughly $620/(1888/2) = 66\%$ of the BAT-detected sources have potential eROSITA counterparts.

For ten newly-detected BAT sources that do not have confirmed counterparts, new eROSITA sources are found within the 12 arcmin error circle and may be the potential counterpart. However, with eROSITA detection alone and without other information about the physical nature of these sources, it is difficult to determine whether the eROSITA source truly associates with the BAT source. Therefore, we do not list the eROSITA source as the official counterpart. 
The only exception is 1eRASS J031010.7-573040, which is only 1.44 arcsec away from the only AGN, ESO 116-10, within the BAT error circle of SWIFT J0310.3-5730. Therefore, it is likely that 1eRASS J031010.7-573040 is the counterpart of ESO 116-10, and thus is associated with SWIFT J0310.3-5730.

\section{Conclusion}
The {\it Swift}-BAT 157 month survey catalog is the sixth catalog of the BAT all-sky hard X-ray survey. This catalog utilizes survey data from the beginning of the {\it Swift} mission, December 2004, to December 2017. The analysis in this catalog reprocesses data from 2007 to include updated gain calibration and pattern noise calculations.

The 157-month survey catalog reaches a sensitivity of $8.83 \times 10^{-12} \rm \ erg \ s^{-1} \ cm^{-2}$ for 90\% of the sky and $6.44 \times 10^{-12} \rm \ erg \ s^{-1} \ cm^{-2}$ for 10\% of the sky. This flux limit is higher than what was expected when only considering statistical noise, indicating the emergence of systematic noise.

In summary, the catalog includes 1888 sources, with 256 newly-detected sources. Similar to previous catalogs, the majority of our new detections with identified counterparts are AGNs, mostly Seyfert I galaxies. We present the characteristics of all the 1888 sources, including the source locations, spectral fits, fluxes, counterpart associations, redshifts and luminosities when the information is available. 

In addition, the spectra, the Crab-weighted monthly lightcurves, the eight-band monthly lightcurves, the eight-band snapshot lightcurves, and quick-look plots are available on the BAT 157-month web page \url{https://swift.gsfc.nasa.gov/results/bs157mon/}.

\section{Future outlook of BAT survey catalog}
Since the early time of the {\it Swift} mission, the BAT team has been publishing the BAT survey catalog once every few years. 
As we enter the time domain and multi-messenger era, we wish to improve the latency of the BAT survey results. Based on previous experience, the most time-consuming parts were usually counterpart identification and investigation of any changes in the instrumental behaviors, especially in recent years. 

The 157-month survey catalog includes data prior to December 2017. At the present time of 2024, BAT has collected another 6.5 years of survey data. During this time, BAT has experienced some instrument issues, including several spontaneous reboots that resulted in additional downtime of observation, problems updating the number of enabled detectors in the downlinked data for many months in 2019, and energy calibration issue for some detectors starting in 2019. 
Moving forward, the BAT survey analysis will need to exclude the additional downtime, implementing manual update of the correct number of the enabled detectors for those months in 2019, and either correct the energy calibration issue or exclude those problematic detectors in the analysis. Once these issues are resolved, we plan to start processing the new survey data.

We plan to implement a new processing pipeline that will continuously analyze new survey data and produce regular updates (e.g., once a month) on the BAT survey web page. We plan to include updates of light curves and spectra for all the currently-detected sources. 
This new interface will enable the community to access the most updated BAT survey results and perform timely counterpart cross-checks with other observations.



{\it Acknowledgement}
We acknowledge the long-term support from the Neil Gehrels {\it Swift} Observatory for conducting this work. This study utilizes the SIMBAD, NED\footnote{The NASA/IPAC Extragalactic Database (NED)
is operated by the Jet Propulsion Laboratory, California Institute of Technology,
under contract with the National Aeronautics and Space Administration.}, and HEASARC online database, as well as the online tools for the XRT data analysis provided by by the UK Swift Science Data Centre at the University of Leicester. We appreciate the great support of the teams that developed and maintain these database and tools. In addition, we thank Michael Moss for valuable input regarding the BAT sensitivity changes due to the decreasing number of enabled detectors. Furthermore, we are grateful for the help and support from J. D. Myers for creating and maintaining the public BAT 157-month web page. In addition, We are grateful of the helpful suggestions from the referee, which has greatly improve the paper.

This material is based upon work supported by the National Aeronautics and Space Administration under Agreement No. 80NSSC23K0552 issued through the Office of Science.  In accordance with Federal law, New Mexico Consortium is prohibited from discriminating on the basis of race, color, national origin, sex, age, or disability.

LM acknowledges that support for this work was provided by NASA through the
NASA Hubble Fellowship grant No. HST-HF2-51486.001-A
awarded by the Space Telescope Science Institute, which is
operated by the Association of Universities for Research in
Astronomy, Inc., for NASA, under contract NAS5-26555.

KO acknowledges support from the Korea Astronomy and Space Science Institute under the R\&D program (Project No. 2025-1-831-01), supervised by the Korea AeroSpace Administration, and the National Research Foundation of Korea (NRF) grant funded by the Korea government (MSIT) (RS-2025-00553982).


Finally, we are grateful for the helpful comments from the anonymous referee that has greatly improved this paper.

\newpage

\setlength{\tabcolsep}{2pt} 
\renewcommand{\arraystretch}{1} 
\begin{rotatetable*}
\begin{deluxetable*}
{ccccccccccccccccccc}
\tabletypesize{\tiny}
\tablecolumns{19} 
\tablewidth{0pt}
\tablecaption{\footnotesize Summary of the sources presented in the 157-month survey catalog (truncated version). A complete table can be found on the 157-month survey catalog web page.}
\tablehead{\colhead{Num} & \colhead{BAT name$^{\rm a}$} & \colhead{RA$^{\rm b}$} & \colhead{Dec$^{\rm b}$} & 
\colhead{SNR}  & \colhead{Ctpt name} & \colhead{Ctpt RA$^{\rm c}$} & \colhead{Ctpt Dec$^{\rm c}$} & \colhead{flux$^{\rm d}$} & \colhead{$f_{\rm err}$$^{\rm e}$} & \colhead{$\Gamma$$^{\rm f}$} & \colhead{$\Gamma_{\rm err}$$^{\rm g}$} & \colhead{$\chi^2_{r}$} & \colhead{z$^{\rm h}$} & \colhead{Lum$^{\rm i}$} & \colhead{C$^{\rm j}$} & \colhead{AS$^{\rm k}$} & \colhead{Class$^{\rm l}$} & \colhead{Type}} 
\startdata
1 &    SWIFT J0001.0-0708 & 0.220 & -7.140 & 9.02 & 2MASX J00004876-0709117 & 0.2032 & -7.1532 & 15.63 & 12.78-18.75 & 2.17 & 1.92-2.44 & 1.85 & 0.0375 & 43.72 &    & 1 & 5 & Sy1.9  \\
2 &    SWIFT J0001.6-7701 & 0.364 & -77.005 & 5.30 & Fairall 1203 & 0.4419 & -76.9540 & 9.77 & 7.06-12.75 & 1.88 & 1.51-2.30 & 3.15 & 0.0584 & 43.91 &    & 1 & 4 & Sy1  \\
3 &    SWIFT J0002.5+0323 & 0.614 & 3.372 & 6.11 & NGC 7811 & 0.6103 & 3.3519 & 11.93 & 8.44-15.77 & 1.75 & 1.35-2.20 & 0.88 & 0.0255 & 43.26 &    & 1 & 4 & Sy1.5  \\
4 &    SWIFT J0003.3+2737 & 0.880 & 27.632 & 6.54 & 2MASX J00032742+2739173 & 0.8643 & 27.6548 & 11.52 & 8.59-14.77 & 1.98 & 1.61-2.38 & 0.54 & 0.0396 & 43.64 &    & 1 & 5 & Sy2  \\
5 &    SWIFT J0005.0+7021 & 0.982 & 70.332 & 7.10 & 2MASX J00040192+7019185 & 1.0082 & 70.3217 & 12.26 & 9.45-15.32 & 1.88 & 1.56-2.23 & 0.97 & 0.0960 & 44.47 &    & 1 & 5 & Sy1.9  \\
6 &    SWIFT J0006.2+2012 & 1.570 & 20.228 & 10.01 & Mrk 335 & 1.5813 & 20.2029 & 13.92 & 11.43-16.64 & 2.35 & 2.09-2.65 & 1.83 & 0.0258 & 43.34 &    & 1 & 4 & Sy1.2  \\
7 &    SWIFT J0009.4-0037 & 2.321 & -0.628 & 5.66 & 2MASX J00091156-0036551 & 2.2982 & -0.6152 & 10.57 & 7.18-14.40 & 1.76 & 1.32-2.25 & 0.57 & 0.0733 & 44.15 &    & 1 & 5 & Sy2  \\
8 &    SWIFT J0010.5+1057 & 2.613 & 10.968 & 14.79 & Mrk 1501 & 2.6292 & 10.9749 & 27.27 & 23.99-30.72 & 1.84 & 1.68-2.01 & 1.33 & 0.0893 & 44.75 &    & 1 & 7 & BZQ  \\
9 &    SWIFT J0017.1+8134 & 4.625 & 81.565 & 8.69 & [HB89] 0014+813 & 4.2853 & 81.5856 & 11.28 & 8.72-14.12 & 2.18 & 1.82-2.61 & 2.41 & 3.3660 & 48.07 &    & 1 & 7 & BZQ  \\
10 &    SWIFT J0021.2-1909 & 5.289 & -19.169 & 10.38 & 2MASX J00210753-1910056 & 5.2814 & -19.1682 & 16.25 & 13.37-19.36 & 2.18 & 1.93-2.47 & 0.20 & 0.0956 & 44.59 &    & 1 & 5 & Sy1.9  \\
11 &    SWIFT J0023.2+6142 & 5.818 & 61.663 & 8.19 & IGR J00234+6141 & 5.7000 & 61.6600 & 11.33 & 8.95-13.95 & 2.29 & 1.96-2.66 & 0.46 &      NA  &         NA &    & 1 & 9 & CV  \\
12 &    SWIFT J0025.2+6410 & 6.302 & 64.140 & 11.93 & Tycho SNR & 6.2840 & 64.1650 & 12.82 & 10.97-14.86 & 3.06 & 2.74-3.43 & 2.37 &      NA  &         NA &    & 1 & 11 & SNR  \\
13 &    SWIFT J0025.8+6818 & 6.401 & 68.389 & 9.83 & 2MASX J00253292+6821442 & 6.3870 & 68.3623 & 18.71 & 15.56-22.05 & 1.64 & 1.40-1.89 & 0.40 & 0.0120 & 42.79 &    & 1 & 5 & Sy2  \\
14 &    SWIFT J0026.5-5308 & 6.692 & -53.167 & 8.70 & 2MASX J00264073-5309479 & 6.6695 & -53.1633 & 13.59 & 10.69-16.72 & 1.82 & 1.52-2.14 & 1.30 & 0.0629 & 44.12 &    & 1 & 4 & Sy1  \\
15 &    SWIFT J0028.9+5917 & 7.182 & 59.269 & 58.84 & V709 Cas & 7.2036 & 59.2894 & 76.77 & 74.64-78.94 & 2.57 & 2.52-2.62 & 18.18 &      NA  &         NA &    & 1 & 9 & CV  \\
16 &    SWIFT J0029.2+1319 & 7.286 & 13.262 & 7.78 & [HB89] 0026+129 & 7.3067 & 13.2675 & 11.03 & 8.60-13.75 & 2.37 & 2.04-2.76 & 0.66 & 0.1420 & 44.79 &    & 1 & 4 & Sy1.2  \\
17 &    SWIFT J0030.0-5904 & 7.683 & -59.008 & 7.95 & ESO112-6 & 7.6826 & -59.0072 & 14.78 & 11.72-18.03 & 1.56 & 1.28-1.84 & 0.75 & 0.0290 & 43.47 &    & 1 & 5 & Sy2  \\
18 &    SWIFT J0033.6+6127 & 8.329 & 61.490 & 10.04 & 2MASX J00331831+6127433 & 8.3266 & 61.4620 & 15.12 & 12.61-17.82 & 2.16 & 1.91-2.42 & 0.83 & 0.1050 & 44.64 &    & 1 & 5 & Sy1.9  \\
19 &    SWIFT J0034.5-7904 & 8.868 & -79.116 & 5.71 & 2MASX J00341665-7905204 & 8.5697 & -79.0890 & 7.61 & 5.55-9.97 & 2.52 & 2.09-3.07 & 0.50 & 0.0740 & 44.02 &    & 1 & 4 & Sy1  \\
20 &    SWIFT J0034.6-0422 & 8.629 & -4.424 & 7.91 & 2MASX J00343284-0424117 & 8.6368 & -4.4033 & 16.27 & 12.87-19.94 & 1.74 & 1.46-2.04 & 0.39 & 0.2130 & 45.35 &    & 1 & 5 & Sy2  \\
21 &    SWIFT J0036.0+5951 & 8.983 & 59.839 & 39.05 & 1ES 0033+595 & 8.9694 & 59.8346 & 50.03 & 47.80-52.32 & 2.55 & 2.48-2.63 & 0.33 & 0.0860 & 44.98 &    & 1 & 7 & BZB  \\
22 &    SWIFT J0036.3+4540 & 9.084 & 45.680 & 11.38 & CGCG 535-012 & 9.0874 & 45.6650 & 19.19 & 16.30-22.25 & 1.83 & 1.62-2.05 & 1.38 & 0.0476 & 44.02 &    & 1 & 4 & Sy1.2  \\
23 &    SWIFT J0037.2+6123 & 9.260 & 61.365 & 11.88 & BD +60 73 & 9.2902 & 61.3601 & 16.97 & 14.66-19.43 & 2.36 & 2.15-2.60 & 2.21 &      NA  &         NA &    & 1 & 13 & HMXB  \\
24 &    SWIFT J0038.4+2337 & 9.653 & 23.594 & 9.41 & Mrk 344 & 9.6339 & 23.6133 & 15.19 & 12.17-18.47 & 1.93 & 1.65-2.24 & 0.51 & 0.0249 & 43.34 &    & 1 & 5 & Sy2  \\
25 &    SWIFT J0041.0+2444 & 10.220 & 24.788 & 3.97 & WISEAJ004039.88+244539.3 & 10.1662 & 24.7609 & 5.28 & 3.02-8.15 & 2.41 & 1.72-3.39 & 0.49 & 0.0784 & 43.91 &    & 1 & 5 & Sy1.9  \\
26 &    SWIFT J0041.9-0921 & 10.449 & -9.344 & 6.58 & ABELL 85 & 10.4075 & -9.3425 & 7.22 & 5.88-8.75 & 4.35 & 3.67-5.25 & 0.26 &      NA  &         NA &    & 1 & 3 & Galaxy Cluster  \\
27 &    SWIFT J0042.6+4112 & 10.679 & 41.220 & 8.30 & SWIFT J0042.7+4111 & 10.6679 & 41.2002 & 8.92 & 7.26-10.79 & 3.19 & 2.78-3.69 & 1.43 &      NA  &         NA &    & 1 & 10 & Pulsar  \\
28 &    SWIFT J0042.9-2332 & 10.714 & -23.537 & 27.06 & NGC 235A & 10.7200 & -23.5410 & 47.25 & 44.04-50.57 & 1.85 & 1.75-1.94 & 4.20 & 0.0222 & 43.74 &    & 1 & 5 & Sy1.9  \\
29 &   SWIFT J0042.9+3016A & 10.809 & 30.296 & 10.45 & 2MASX J00430184+3017195 & 10.7578 & 30.2888 & 4.65 & 2.33-6.96 & 2.14 & 1.63-3.21 & 2.23 & 0.0489 & 43.43 &  Y  & 1 & 5 & Sy2  \\
30 &   SWIFT J0042.9+3016B & 10.729 & 30.292 & 11.12 & 2MASX J00423991+3017515 & 10.6663 & 30.2976 & 13.76 & 10.83-16.93 & 2.17 & 1.84-2.57 & 1.91 & 0.1408 & 44.88 &  Y  & 1 & 7 & BZQ  \\
31 &    SWIFT J0042.9-1135 & 10.769 & -11.594 & 5.20 & MCG -02-02-095 & 10.7865 & -11.6010 & 8.60 & 5.72-11.93 & 2.12 & 1.63-2.73 & 0.57 & 0.0189 & 42.85 &    & 1 & 5 & Sy2  \\
32 &    SWIFT J0046.2-4008 & 11.556 & -40.095 & 8.59 & ESP 39607 & 11.5860 & -40.0970 & 12.87 & 10.21-15.77 & 1.98 & 1.70-2.29 & 2.29 & 0.2013 & 45.19 &    & 1 & 5 & Sy2  \\
33 &    SWIFT J0048.8+3155 & 12.195 & 31.957 & 87.68 & Mrk 348 & 12.1964 & 31.9570 & 154.75 & 151.67-157.85 & 1.83 & 1.80-1.86 & 10.57 & 0.0150 & 43.91 &    & 1 & 5 & Sy1.9  \\
34 &    SWIFT J0051.6+2928 & 12.884 & 29.469 & 4.96 & MCG +05-03-013 & 12.8959 & 29.4013 & 7.45 & 4.68-10.75 & 2.13 & 1.55-2.91 & 0.72 & 0.0360 & 43.36 &    & 1 & 4 & Sy1  \\
35 &    SWIFT J0051.8-7320 & 12.594 & -73.194 & 9.64 & RX J0052.1-7319 & 13.0921 & -73.3181 & 13.45 & 11.06-16.05 & 2.33 & 2.06-2.64 & 2.26 &      NA  &         NA &    & 1 & 13 & HMXB  \\
36 &    SWIFT J0051.9+1724 & 12.961 & 17.423 & 21.18 & Mrk 1148 & 12.9783 & 17.4329 & 29.08 & 26.37-31.92 & 2.28 & 2.14-2.43 & 1.16 & 0.0640 & 44.47 &    & 1 & 4 & Sy1.5  \\
37 &    SWIFT J0052.3-2730 & 13.062 & -27.495 & 3.99 & 2MASX J00520383-2723488 & 13.0158 & -27.3969 & 6.90 & 4.17-10.14 & 1.93 & 1.37-2.56 & 2.08 & 0.0768 & 44.01 &    & 1 & 5 & Sy2  \\
38 &    SWIFT J0053.3-7224 & 13.732 & -72.475 & 13.23 & RX J0053.8-7226 & 13.4790 & -72.4460 & 15.63 & 13.72-17.68 & 2.96 & 2.71-3.24 & 0.72 &      NA  &         NA &    & 1 & 13 & HMXB  \\
39 &    SWIFT J0054.9+2524 & 13.705 & 25.451 & 8.14 & [HB89] 0052+251 & 13.7171 & 25.4272 & 14.96 & 11.89-18.30 & 1.79 & 1.51-2.07 & 1.33 & 0.1550 & 45.00 &    & 1 & 4 & Sy1.2  \\
40 &    SWIFT J0055.4+4612 & 13.836 & 46.214 & 19.71 & 1RXS J005528.0+461143 & 13.8329 & 46.2158 & 21.83 & 20.05-23.70 & 2.97 & 2.81-3.15 & 2.71 &      NA  &         NA &    & 1 & 9 & CV  \\
\enddata
\tablecomments{$^{\rm a}$ BAT name. For sources existed in previous catalogs, we keep the same name. For new sources, the name is derived from the BAT position (J2000 coordinate). 
\\ $^{\rm b}$ The BAT source positions found in the 157-month catalog. \\
$^{\rm c}$ Counterpart RA and Dec, mostly from either SIMBAD or NED. \\
$^{\rm d}$ The BAT flux ($14-195$ keV) estimated from the BAT mosaic maps using the BAT positions. The unit is in $10^{-12} \, \rm erg \, s^{-1} \, cm^{-2}$ \\
$^{\rm e}$ The flux error range is the 90\% confidence interval.\\
$^{\rm f}$ The spectral index is computed from the power-law fit to the eight-band data. \\
$^{\rm g}$ The error range to the spectral index, which represents the 90\% confidence interval. \\
$^{\rm h}$ The redshift is taken from SIMBAD or NED. \\
$^{\rm i}$ The luminosity that corresponds to the observed $14-195$ keV band. The luminosity is computed from the flux and redshift in this table, with units of $\rm log(erg/s)$. The calculation assumes $H_0 = 69 {\rm km \, s^{-1} \, Mpc^{-1}}, \Omega_m = 0.3, \Omega_{\Lambda}=0.7$. \\
$^{\rm j}$ ``Y'' indicates potential contamination from nearby sources. \\
$^{\rm k}$ Associate strength. See description in Sect.~\ref{sect:data_analysis} and Table~\ref{tab:ctpt_stren}. \\
$^{\rm l}$ Source class. See the definition in Table~\ref{tab:source_type}. \\
References: \citet{Oh18}, \citet{Wenger00_SIMBAD}, \citet{Helou91_NED}, \citet{2SXPS}, \citet{Pavlinsky21_ARTXC, Merloni24_eROSITA24}
\citet{Pablo23}, \citet{Fortin18}, \citet{Yukita17}, \citet{deMartino20}, \citet{Fortin23}, \citet{Fortin24}, \citet{Gaia22}, \citet{Galloway08}, \citet{Galloway20}, \citet{Giacconi62}, \citet{Gomez21}, \citet{Harris90}, \citet{Koss22}, \citet{Halpern18}, \citet{Liu01}, \citet{Kennedy20}, \citet{Markwardt08}, \citet{Halpern18}, \citet{Szkody20}
}
\end{deluxetable*}
\end{rotatetable*}


\appendix
\subsection{Detailed update of the gain calibration}
\label{sect:gain_appendix}

As noted in Sect.~\ref{sect:gain}, the BAT energy scale gradually changes due to
radiation damage within the CZT detectors.  The primary impact of these changes is a downward shift
in the overall gain scale factor, in keV per analog to data unit (ADU) units.  Such
effects have been previously noted for similar CdTe gamma-ray detectors on board {\it INTEGRAL}'s ISGRI
detector systems \citep{Lebrun96,Lebrun05}, both in pre-flight testing and in in-flight performance,
and in other irradiated CZT detectors \citep{Shy23}.
This type of degradation is presumably due to detector irradiation by high energy cosmic rays and 
trapped protons, which damage the CZT crystalline lattice, resulting in
increased charge trapping and reduced charge collection.  Such degradation also leads to a small and 
gradual degradation in the off-diagonal portion of the response (i.e., the resolution), which we do not
treat further.

For the BAT survey, X-ray photons are detected and binned into 80 bins using a nominal ADU to energy conversion 
which was estimated before launch.  This conversion does not account for any post-launch degradation.
Thus, counts of a fixed energy will gradually migrate to lower apparent energy bins.  The effect
is different in each BAT detector.  This drift leads to two undesireable effects.  The  
histograms bin edges will have incorrectly assigned energies by default, which will lead to 
the incorrect spectrum of astrophysical sources.  In addition, since the drift is different for
each detector, the shifted counts are distributed unevenly across the detector array, leading to increased
image coding noise.  We correct each DPH using the following process.

BAT has two $^{241}$Am ``tagged'' sources which illuminate the array.  The $\alpha$ decay
of the $^{241}$Am nucleus produces a series of decay lines, the primary and 
cleanest of which is at an energy 59.5 keV.  The BAT $^{241}$Am
sources use detection of the coincident $\alpha$ particle \citep{Barthelmy05} to tag events as calibration 
events, and dedicated per-detector calibration spectra are accumulated on-board the BAT.  We further accumulate 
tagged spectra for each detector on six month timescales, and fit the centroid of the expected 59.5 keV line.
Survey spectra are adjusted by shifting counts into their correct energy bins, again on a detector-by-detector
basis, using the {\it baterebin} tool.  The resulting BAT images and spectra have the correct energy scale.

\begin{figure}[!h]
\begin{center}
\includegraphics[width=0.5\textwidth]{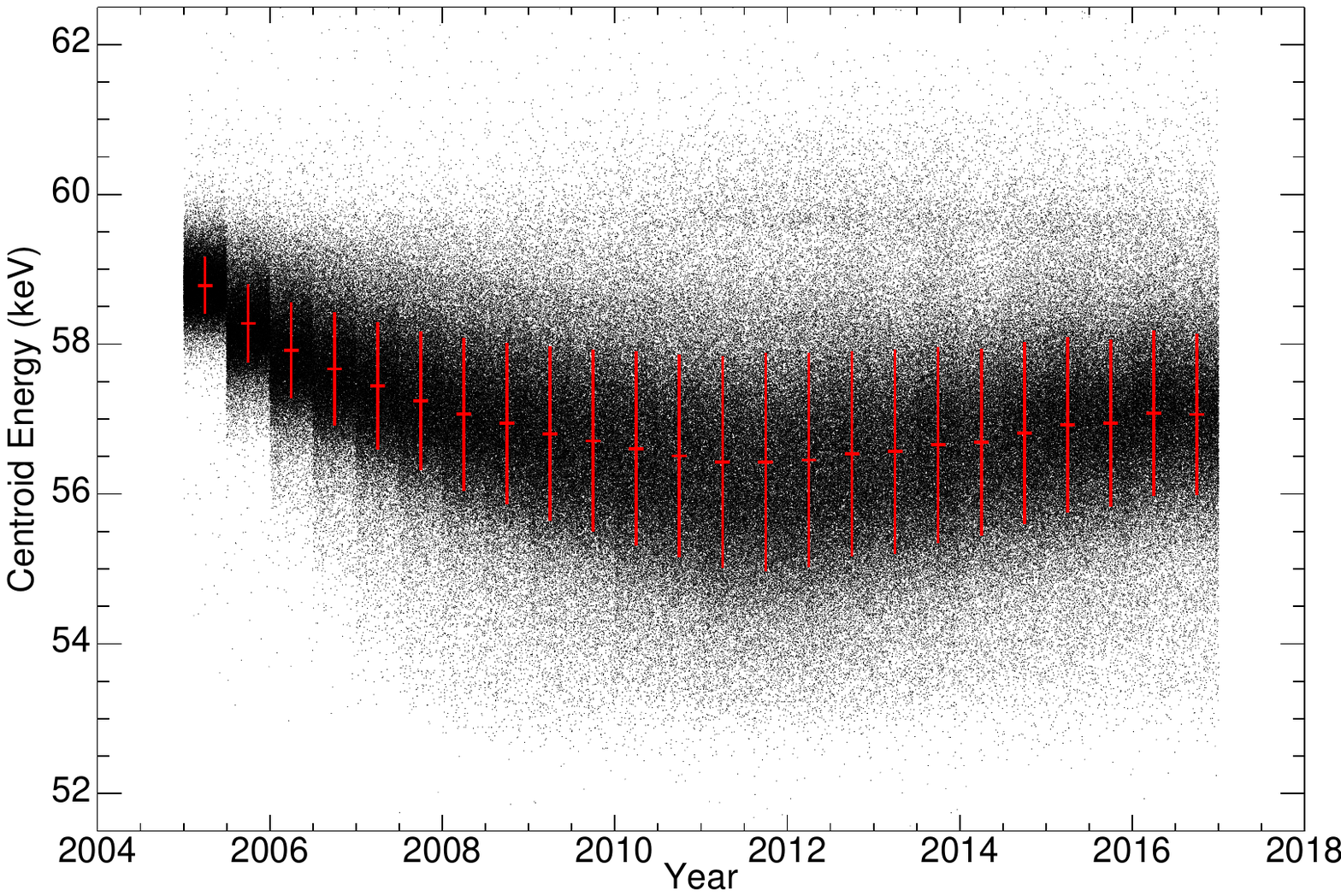}
\end{center}
\caption{
{\it Swift} BAT gain variations over 2004-2017 time period, shown
as the fitted centroid of the 59.5~keV $^{241}$Am gamma-ray line.
BAT tagged spectra were accumulated at six month intervals and the
line centroid was fitted for each detector and displayed.  On the
X-axis, a $\pm$0.5~yr random offset was applied to each detector for clarity of display.
The array mean and standard deviation for each six month interval is shown in red.
\label{fig:BAT_gain}
}
\end{figure}

The on-orbit performance of the BAT energy scale is shown in Figure~\ref{fig:BAT_gain} for launch through 2016.
The figure shows the fitted centroid of the 59.5 keV $^{241}$Am line versus time, and by detector.  Through
year $\sim$2010, the gain scale shifted by about 1\% yr$^{-1}$ ($\sim$3 keV over five years).  After that
year, the array-averaged gain scale actually recovered somewhat by about 1 keV, although individual
detectors have increased scatter from the mean.  During the years 2005-2010, the {\it Swift} spacecraft
was at a relatively high altitude of 585--595~km, and over the 2010-2018 timescale the altitude decreased 
significantly to about 550 km, primarily due to increased atmospheric drag related to solar cycle 24.
As the energetic proton density in the SAA region decreases by a factor of $\sim$2--3 over that altitude range 
\citep{Fuerst09}, the degradation processes due to radiation damage appear to have been significantly
curtailed at the lower altitudes.  

The slight gain recovery seen after 2010 is likely not a mistake.  Some radiation damage is known to "heal" over time, 
especially at higher temperatures, which enhances the mobility of crystalline defects \citep{Fraboni05}.  
BAT is operated near room temperature, which is favorable to the recovery process.



\bibliographystyle{apj}
\bibliography{ref}

\end{document}